\documentclass[submission,copyright,creativecommons]{eptcs}
\providecommand{\event}{PDMC 2011} 
\usepackage{breakurl}              

\title{Variations on Multi-Core Nested Depth-First Search}
\author{Alfons Laarman \qquad\qquad Jaco van de Pol
\institute{Formal Methods and Tools, University of Twente, The Netherlands}
\email{\{a.w.laarman,vdpol\}@cs.utwente.nl}
}
\def\titlerunning{Variations on Multi-Core NDFS}
\def\authorrunning{A.W. Laarman and J.C. van de Pol}

\usepackage{tikz}
\usepackage{listings}
\usepackage{multicol}
\usepackage{multirow}
\usepackage[T1]{fontenc}
\usepackage{rotating}
\usepackage{wrapfig}

\newtheorem{theorem}{Theorem}
\newtheorem{definition}[theorem]{Definition}

\newcommand{\concept}[1]{\textit{#1}}

\newcommand{\acc}{\ensuremath{\mathcal{A}}}

\newcommand{\buchi}{\ensuremath{\mathcal{B}}}
\newcommand{\post}{\ensuremath{\mathsf{post}}}

\newcommand{\states}{\ensuremath{\mathcal{S}}}

\newcommand{\sint}{\ensuremath{s_I}}

\newcommand{\dfsred}{\textsf{dfs\_red}}

\newcommand{\ndfs}{\textsf{ndfs}}

\newcommand{\DFS}{\mbox{\sc Dfs}}
\newcommand{\DFSs}{\mbox{\sc Dfs}s}
\newcommand{\BFS}{\mbox{\sc Bfs}}
\newcommand{\NDFS}{\mbox{\sc Ndfs}}
\newcommand{\ENDFS}{\mbox{\sc ENdfs}}
\newcommand{\LNDFS}{\mbox{\sc LNdfs}}

\newcommand{\NMNDFS}{\mbox{\sc NMc-ndfs}}
\newcommand{\MAP}{\mbox{\sc Map}}

\newcommand{\OWCTY}{\mbox{\sc Owcty}}
\newcommand{\OWCTYOTF}{\mbox{\sc Owcty-Map}}

\newcommand{\BEEM}{\mbox{\sc Beem}}
\newcommand{\LTSmin}{\mbox{\sc LTSmin}}
\newcommand{\Divine}{\mbox{\sc DiVinE}}

\newcommand{\white}{\mathit{white}}
\newcommand{\blue}{\mathit{blue}}
\newcommand{\cyan}{\mathit{cyan}}
\newcommand{\pink}{\mathit{pink}}
\newcommand{\red}{\mathit{red}}

\newcommand{\dangerous}{\mathit{dangerous}}
\newcommand{\cc}{\mathit{color}}
\newcommand{\cnt}{\mathit{count}}
\newcommand{\allred}{\mathit{allred}}

\lstdefinelanguage{simple}{
	morekeywords={for, all, in, do, proc, if, then, else, false, true,
                report, cycle, exit, no, all, cas, while, await, and},
	sensitive=false,
	morecomment=[l]{//},
	morecomment=[s]{/*}{*/}
}
\begin{document}
\maketitle

\begin{abstract}
Recently, two new parallel algorithms for on-the-fly model checking of
LTL properties were presented at the same
conference: {\em Automated Technology for Verification and Analysis, 2011}. Both
approaches extend Swarmed \NDFS, which runs several sequential \NDFS\ instances in parallel. 
While parallel random search already speeds up detection of bugs, 
the workers must share some global information in order to speed up full 
verification of correct models. The two algorithms
differ considerably in the global information shared between workers,
and in the way they synchronize.

Here, we provide a thorough experimental comparison between the two
algorithms, by measuring the runtime of their implementations on
a multi-core machine. Both algorithms were implemented in the same
framework of the model checker \LTSmin, using similar 
optimizations, and have been subjected to the full \BEEM\ model database.

Because both algorithms have complementary advantages, we
constructed an algorithm that combines both ideas. This combination
clearly has an improved speedup. We also compare the results with the
alternative parallel algorithm for accepting cycle detection 
\OWCTYOTF.
Finally, we study a simple statistical model
for input models that do contain accepting cycles.
The goal is to distinguish the speedup due to parallel random search from
the speedup that can be attributed to clever work sharing schemes.
\end{abstract}


\section{Introduction}

Model checking is an important technique to automatically verify that
a system's behavior is free from subtle bugs, for instance
violations of safety and liveness requirements. 
Linear Time Logic (LTL) expresses such requirements as properties on
individual runs of a system. LTL model checking reduces to detecting
accepting cycles in a so-called B\"uchi
automaton~\cite{AutoVerif1986}. A linear-time algorithm to detect
those cycles is the Nested Depth-First Search (\NDFS) algorithm,
introduced by Courcoubetis et al.~\cite{courcoubetis}. \NDFS\ can also
terminate as soon as some accepting cycle is found, which makes it 
very useful for bug hunting. Still, model
checking is a time- and memory-consuming procedure due to the sheer
size of the state space of realistic systems, leading to an extremely
large B\"uchi automaton.

During the last decades, processor speeds have been greatly increased,
making model checkers much more powerful. Where early papers discussed
the verification of models with a few thousand states, currently we
can easily handle billions of states \cite{ltsmin,boosting,freelunch}.  Recently,
however, these advances are grinding to a halt, because of physical
limits inside the CPU cores. Instead, the number of logical computing
cores increases. Nonetheless, model checking can still benefit from
the progress made by CPU manufacturers, if the algorithms are
parallelized.

A complication is that \DFS\ (and thus \NDFS) is inherently sequential
\cite{reif}.  Barnat et al.\ have therefore introduced breadth-first
search (\BFS) based algorithms, such as {\em Maximal-Accepting-Predecessors}
(\MAP~\cite{BrnoMAP}) and {\it One-Way-Catch-Them-Young}
(\OWCTY~\cite{BrnoOWCTY}). These algorithms deliver excellent speedups,
but sacrifice linear-time complexity. However,
their latest combined \OWCTYOTF\ algorithm \cite{owcty-otf}, is 
linear-time for the class of weak LTL properties and also useful for
bug hunting. It is therefore the current state of the art in
multi-core LTL model checking.

Recently, also two parallel
\NDFS-based algorithms were introduced~\cite{Evangelista11,Laarman11}. 
Both take as starting point a
randomized parallel search by a swarm of \NDFS\ workers. While this is
useful for bug-hunting, it does not really help in the absence of
bugs, in which case all workers traverse the full state space. To
improve speedup, both algorithms share some global information between
workers, in order to reduce the amount of work even in the absence of
accepting cycles.  \ENDFS\ from Evangelista et al.~\cite{Evangelista11} shares a
lot of information, but this may break the required \DFS\ order. A
sequential repair procedure steps in when a potentially dangerous
situation is detected.  On the other hand, \LNDFS\ from Laarman et
al.~\cite{Laarman11} shares less global information and adds extra
synchronization. This avoids dangerous situations and the need for a
repair strategy. However, this leads to a reduced amount of work
sharing in some cases.

\paragraph*{Contributions.} 
The main goal of this paper is to experimentally compare both multi-core
\NDFS\ algorithms. In order to enable a fair comparison, we extended
\ENDFS\ with the same optimizations as used in \LNDFS. We implemented
both algorithms in the same framework of \LTSmin. Finally, we subjected both
implementations to the full \BEEM\ benchmark database~\cite{beem}, running them
on shared memory machines with up to 16 cores. Note that actual runtimes
had not yet been reported for \ENDFS, although workload distributions were
shown in~\cite{Evangelista11}. Also, for \LNDFS, we have rerun the experiments
from~\cite{Laarman11}.

Another contribution is a simple combination of the \ENDFS\ and \LNDFS\ algorithms,
improving the speedup compared to both of them. We also compare all mentioned
algorithms with the \OWCTYOTF\ algorithm, both for bug hunting and for
full verification. Finally, based on a simple statistical 
model~\cite{HyvarinenJN08}, we investigate how much of the speedup
in the parallel \NDFS\ algorithms should be contributed to the effects
of parallel random search and what is the contribution of the more
clever work sharing schemes. 

The algorithms are explained in Section~\ref{s:algorithms}. 
The experimental results are presented in Section~\ref{s:experiments}. 
Section~\ref{s:discussion} contains the discussion on parallel random search.
Our conclusions are summarized in Section~\ref{s:conclusion}.


\section{Parallel Algorithms to Detect Accepting Cycles}
\label{s:algorithms}

Model checking properties from Linear Temporal Logic (LTL) entails
verifying that all runs of a given system satisfy some safety or liveness
property. In the automata-theoretic approach~\cite{AutoVerif1986,principles.mc}, 
a B\"uchi automaton is constructed that accepts all infinite words
corresponding to those runs of the original system that violate the
property. So the problem is reduced to the emptiness check of
$\omega$-regular languages. A B\"uchi automaton accepts a word if it
visits some accepting state infinitely often. For finite automata,
this implies that there is a cycle through some accepting state.

\begin{definition}
A \emph{B\"uchi automaton} is a quadruple $\buchi = (\states,
\sint, \post, \acc)$, where $\states$ is the finite set of states, $\sint\in\states$
is the initial state, $\post:\states\to 2^\states$ the successor
function, and $\acc\subseteq\states$ the set of accepting states.
\end{definition}

Note, that the use of the \post\ function reflects the way in which
the B\"uchi automaton is computed \concept{on-the-fly} from the input
model. When appropriate, we refer to the complete automaton as graph
or state~space.

The purpose of all algorithms in this paper is to detect an accepting
cycle in this graph. For states $s,t\in\states$, we write $s\rightarrow t$ if
$t\in\post(s)$, and $\rightarrow^+$ ($\rightarrow^*$), for its
(reflexive) transitive closure. An accepting cycle is some state
$a\in\acc$, which is reachable from the initial state
($\sint\rightarrow^* a$) and lies on a non-trivial cycle ($a
\rightarrow^+ a$).

\subsection{Nested Depth-First Search}

The first linear-time algorithm to detect accepting cycles was proposed
by Courcoubetis et al.~\cite{courcoubetis} and is referred to as
Nested Depth-First Search (\NDFS). \NDFS\ also enjoys the on-the-fly
property. This means that the algorithm can terminate as soon as a
cycle is detected, without the need to visit (or even construct) the
whole graph. This makes \NDFS\ very suitable for bug hunting, besides
its use for full verification. Various extensions and optimizations 
to \NDFS\ have been proposed~\cite{holzmann-ndfs,schwoon,gaiser}.
Alg.~\ref{alg:ndfs} most closely resembles 
\textit{New \NDFS}~\cite{schwoon}.

\begin{figure}[t!]
\hspace{0.3cm}
\begin{minipage}[b]{14cm}
\begin{multicols}{2}
\lstset{numbers=left, numberstyle=\tiny}
\begin{lstlisting}
proc ndfs$(s)$								(*\label{nn:ndfs}*)
	dfs_blue$(s)$							(*\label{nn:callblue1}*)
	report no cycle				(*\vspace{2ex}\label{nn:nocycle}*)
proc dfs_red$(s)$							(*\label{nn:dfsred}*)
	$s.\cc$ := $\red$						(*\label{nn:red2}*)
	for all $t$ in $\post(s)$ do			(*\label{nn:forred}*)
		if $t.\cc$ = $\cyan$				(*\label{nn:ifcyan}*)
			report cycle & exit				(*\label{nn:report}*)
		else if $t.\cc$ = $\blue$			(*\label{nn:ifblue}*)
			dfs_red$(t)$					(*\label{nn:callred2}*)
\end{lstlisting}
	\columnbreak
\begin{lstlisting}[firstnumber=last]
proc dfs_blue$(s)$							(*\label{nn:dfsblue}*)
	$s.\cc$ := $\cyan$						(*\label{nn:cyan}*)
	for all $t$ in $\post(s)$ do			(*\label{nn:forblue}*)
		if $t.\cc$ = $\cyan$ and ($s\in\acc \vee t\in\acc$)(*\label{nn:earlydetect1}*)
			report cycle & exit				(*\label{nn:earlydetect2}*)
		if $t.\cc$ = $\white$				(*\label{nn:ifwhite}*)
			dfs_blue$(t)$			 		(*\label{nn:callblue2}*)
	if $s \in\acc$							(*\label{nn:ifacc}*)
		dfs_red$(s)$						(*\label{nn:callred1}*)
	else									(*\label{nn:elseacc}*)
		$s.\cc$ := $\blue$					(*\label{nn:blue}*)
\end{lstlisting}
\end{multicols}
\end{minipage}
\caption{The (sequential) New \NDFS\ algorithm adapted from~\cite{schwoon}}
\label{alg:ndfs}
\end{figure}

In Alg.~\ref{alg:ndfs}, \ndfs($s_I$) initiates a {\em blue \DFS}
from the initial state, so called since explored states are colored
blue (we assume that initially all states are white). A newly visited
state is first colored {\em cyan} (``it is on the DFS-stack''), and
during backtracking after exploration, it is colored full {\em blue}.
However, if at l.\ref{nn:ifacc} the blue \DFS\ backtracks over an
accepting state $s\in \acc$, then $\dfsred(s)$ is called, which is the
nested {\em red} \DFS\ to determine whether there exists a cycle
containing $s$. As soon as a cyan state is found on l.\ref{nn:ifcyan},
an accepting cycle is reported \cite{holzmann-ndfs,schwoon}.
\concept{Early cycle detection}
is also possible in the blue \DFS\ at
l.\ref{nn:earlydetect1},\ref{nn:earlydetect2}. Due to early
cycle detection, it does not matter that the cyan color of $s$ is
overwritten by red at~l.\ref{nn:red2} \cite[Sect.~4.4]{Laarman11}.

\NDFS\ runs in linear time, since each reachable state is visited at
most twice, once in the blue \DFS\ and once in a red \DFS. The
correctness of \NDFS\ essentially depends on the fact that the red
\DFSs\ are initiated on accepting states in the post order imposed by
the blue \DFS. So the red search will never hit another accepting
state that is not already red. 

\subsection{Embarrassing Parallelization: Swarmed N{\small DFS}}

The inherently DFS nature of the blue search makes \NDFS\ hard to
parallelize, since computing the post order is a P-complete
problem~\cite{reif}. One response has been to develop entirely
different algorithms based on Breadth-First Search, cf.\ Sec.~\ref{owcty-map}.

Another approach would be to simply run $N$ isolated instances of \NDFS\ 
(Alg.~\ref{alg:ndfs}) in parallel, in the hope that this \concept{swarm} 
of \NDFS\ workers will detect accepting cycles
earlier~\cite{holzmann-swarm,Laarman11}. Local permutations of the \post\ 
function direct the workers to different regions of the state space,
so their search becomes independent. With $\post^b_i$ ($\post^r_i$) we
denote the permutation of successors used in the blue (red) \DFS\ by
worker $i$. Section~\ref{s:discussion} analyses the expected and actual
improvements due to parallel randomized search.

Although Swarmed \NDFS\ is expected to be profitable for bug hunting, it
does not show a speedup in the absence of accepting cycles, in which
case all workers have to go through the complete state space. Indeed, the
worst-case complexity of all parallel \NDFS\ variations in this paper
is ${\cal O}(|\rightarrow|\cdot|N|)$, i.e. linear both in the size of the 
B\"uchi automaton and in the number of workers.

In order to improve average speedup, some more synchronization between
the workers is needed. Note that a naive global sharing of colors
between multiple workers would be incorrect, because it would destroy
the post-order properties on which \NDFS\ relies. Next, we discuss two
recent proposals for sharing information between the \NDFS\ workers.

\subsection{LN{\small DFS}: Sharing the Red Color Globally}
\label{s:lndfs}

\begin{figure}[t!]
\hspace{0.3cm}
\begin{minipage}[b]{14cm}
\begin{multicols}{2}
\lstset{numbers=left, numberstyle=\tiny}
\begin{lstlisting}
proc lndfs$(s,N)$								(*\label{l:pndfs}*)
	dfs_blue$(s,1)\|..\|$dfs_blue$(s,N)$		(*\label{l:callblue}*)
	report no cycle					(*\vspace{2ex}\label{l:nocycle}*)
proc dfs_red$(s,i)$								(*\label{l:dfsred}*)
	$s.\cc[i]$ := $\pink$      					(*\label{l:pink}*)
	for all t in $\post^r_i(s)$ do    			(*\label{l:forred}*)
		if $t.\cc[i]=\cyan$ 					(*\label{l:ifcyan}*)
			report cycle & exit all				(*\label{l:report}*)
		if $t.\cc[i]\neq\pink\land\neg t.\red$	(*\label{l:ifnred}*)
			dfs_red$(t,i)$						(*\label{l:callred2}*)
	if $s\in\acc$								(*\label{l:ifacc}*)
		$s.\cnt$ := $s.\cnt - 1$
		await $s.\cnt=0$ 						(*\label{l:wait}*)
	$s.\red$ := true							(*\label{l:red2}*)
\end{lstlisting}
	\columnbreak
\begin{lstlisting}[firstnumber=last]
proc dfs_blue$(s,i)$							(*\label{l:dfsblue}*)
	$\allred$ := true							(*\label{l:allred1}*)
	$s.\cc[i]$ := $\cyan$						(*\label{l:cyan}*)
	for all t in $\post^b_i(s)$ do				(*\label{l:forblue}*)
		if $t.\cc[i]=\cyan$ and $(s\in\acc\lor t\in\acc)$(*\label{l:ecd}*)
			report cycle & exit all				(*\label{l:report2}*)
		if $t.\cc[i]=\white\land\neg t.\red$	(*\label{l:ifnew}*)
			dfs_blue$(t,i)$						(*\label{l:callblue2}*)
		if $\neg t.\red$						(*\label{l:allred2}*)
			$\allred$ := false
	if $\allred$								(*\label{l:ifallred}*)
		$s.\red$ := true						(*\label{l:red1}*)
	else if $s\in\acc$	 						(*\label{l:ifred}*)
		$s.\cnt$ := $s.\cnt + 1$
		dfs_red$(s,i)$							(*\label{l:callred1}*)
	$s.\cc[i]$ := $\blue$							(*\label{l:blue}*)
\end{lstlisting}
\end{multicols}
\end{minipage}
\caption{The \LNDFS\ algorithm, pruning blue and red \DFS\ by a global red color, adapted from~\cite{Laarman11}.}
\label{alg:lndfs}
\end{figure}

The basic idea behind \LNDFS\ in Alg.~\ref{alg:lndfs} is to share
information in the backtrack of the red \DFSs\ \cite{Laarman11}. A new
\textit{pink} color is introduced at l.\ref{l:pink} to signify states on the stack of a
red \DFS, analogous to cyan for a blue \DFS. The cyan, blue and pink
colors are all local to worker $i$, but the red color is shared
{\em globally}.  On backtracking from the red \DFS, states are colored red 
at l.\ref{l:red2}. These red states are ignored by {\it
  all blue and red} \DFSs\ (l.\ref{l:ifnew},\ref{l:ifnred}), thus pruning 
the search space for all
workers $i$. To improve pruning during the blue search, the amount of red
states is even increased by the \concept{allred} extension
from~\cite{gaiser} (l.\ref{l:allred1} and l.\ref{l:allred2}-\ref{l:red1}).

To ensure correctness, it is necessary to synchronize the red coloring of 
accepting states (see l.\ref{l:wait}). Otherwise, the algorithm is 
incorrect for more than two workers (see \cite{Laarman11}, which provides 
a correctness proof for $N>0$ workers).
Scalability of the \LNDFS\ algorithm could be hampered by the
need for synchronization, but waiting is only needed when multiple workers
start a red search from the same accepting state; this does not happen 
often in practice. Another reason for limited scalability is that work is 
only pruned when states can be marked red. Despite the allred extension, 
for input graphs with no (or very few) accepting states, all workers 
still have to traverse the whole graph.

\subsection{EN{\small DFS}: an Optimistic Approach with Repair Strategy}

The basic idea of \ENDFS\ in Alg.~\ref{alg:endfs}~\cite{Evangelista11}
is to share both the blue and the red colors globally; only the cyan
and pink colors are local per worker. We deviate from the description
in~\cite{Evangelista11} by adding a cyan stack and
early cycle detection as optimizations,
because this enables a fair comparison with \LNDFS.
Consequently, we also renamed the local colors.

Sharing the blue color can lead to problems, as the post-order is not
preserved by the algorithm. \ENDFS\ optimistically proceeds, but if it
encounters accepting states that are not yet red during the red
search, they are marked dangerous at l.\ref{e:dang}. Eventually,
dangerous states are double-checked in a repair stage, by a separate
sequential \NDFS\ using worker-local colors only, at
l.\ref{e:ifsdang}-\ref{e:lndfs}.  Note that for technical reasons,
states are not colored red during backtracking, but just collected in
the thread-local set $R_i$ at l.\ref{e:accR}. Only after termination
of the red \DFS\ they are made red (provided they are not dangerous) at
l.\ref{e:forR}-\ref{e:red}.

Scalability of the \ENDFS\ algorithm could be hampered by
the repair stage, because this proceeds sequentially. Also,
marking states red occurs relatively late, potentially leading to
more duplicate work within the red \DFS.

\begin{figure}[t!]
\hspace{0.3cm}
\begin{minipage}[b]{14cm}
\begin{multicols}{2}
\lstset{numbers=left, numberstyle=\tiny}
\begin{lstlisting}
proc endfs$(s,N)$									(*\label{e:pndfs}*)
	dfs_blue$(s,1)\|..\|$dfs_blue$(s,N)$			(*\label{e:callblue}*)
	report no cycle						(*\vspace{2ex}\label{e:nocycle}*)
proc dfs_red$(s,i)$									(*\label{e:dfsred}*)
	$s.\pink[i]$ := true   	   						(*\label{e:pink}*)
	$R_i$ := $R_i\cup\{s\}$							(*\label{e:accR}*)
	for all $t$ in $\post_i^r(s)$ do      			(*\label{e:forred}*)
		if $t.\cyan[i]$		 						(*\label{e:ifcyan}*)
			report cycle & exit all					(*\label{e:report}*)
		if $t\in\acc\land\neg t.\red$				(*\label{e:ifacc2}*)
			t.$\dangerous$ := true					(*\label{e:dang}*)
		if $\neg t.\red\land\neg t.\pink[i]$		(*\label{e:ifnred}*)
			dfs_red$(s,i)$							(*\label{e:callred2}*)
\end{lstlisting}
	\columnbreak
\begin{lstlisting}[firstnumber=last]
proc dfs_blue$(s,i)$								(*\label{e:dfsblue}*)
	$s.\cyan[i]$ := true							(*\label{e:bluei}*)
	for all $t$ in $\post_i^b(s)$ do				(*\label{e:forblue}*)
		if $t.\cyan[i]$ and $(s\in\acc\lor t\in\acc)$(*\label{e:ecd}*)
			report cycle & exit all					(*\label{e:report2}*)
		if $\neg t.\cyan[i] \land\neg t.\blue$		(*\label{e:ifnew}*)
			dfs_blue$(t,i)$							(*\label{e:callblue2}*)
	$s.\cyan[i]$ := false							(*\label{e:unbluei}*)
	$s.\blue$ := true								(*\label{e:green}*)
	if s $\in\acc$									(*\label{e:ifacc}*)
		$R_i$ := $\emptyset$						(*\label{e:R}*)
		dfs_red$(s,i)$								(*\label{e:callred1}*)
		for all $r \in R_i$ do						(*\label{e:forR}*)
			if $\neg r.\dangerous\lor s = r	$		(*\label{e:ifrdang}*)
				$r.\red$ := true					(*\label{e:red}*)
		if $s.\dangerous$							(*\label{e:ifsdang}*)
			ndfs$(s, i)$							(*\label{e:lndfs}*)
\end{lstlisting}
\end{multicols}
\end{minipage}
\caption{The optimistic \ENDFS\ algorithm, marking dangerous states, adapted from~\cite{Evangelista11}.}
\label{alg:endfs}
\end{figure}

\subsection{A Combined Version: New M{\small C-NDFS}}
\label{s:nmndfs}

We have recapitulated two very recent multi-core \NDFS\ algorithms, which both
seem to have their merits and pitfalls. \ENDFS, in the end, resorts to a sequential
repair strategy, but it avoids some work duplication due to the global blue color.
\LNDFS\ does not need a repair strategy, but the blue \DFS\ is only pruned when
there are sufficiently many red states, and the algorithm may have to wait 
for synchronization.

A simple idea suggests itself here: we could combine the two algorithms and try
to reconcile their strong points. The idea is simply to run the optimistic
algorithm Alg.~\ref{alg:endfs}, but when dangerous states are encountered at
l.\ref{e:lndfs}, we call the parallel algorithm \LNDFS\ (rather than \NDFS).

We expect an improved speedup, because using \ENDFS\ ensures good work sharing,
even in the absence of accepting states. And using \LNDFS\ parallelizes
the repair strategy, avoiding the important sequential bottleneck of \ENDFS.
In the actual implementation, we also used
a simple load balancing strategy: when a worker finishes \ENDFS,
it starts helping other workers still in their \LNDFS\ repair~phase.

\subsection{One-Way-Catch-Them-Young with Maximal Accepting Predecessors}
\label{owcty-map}

In the next section, we will compare the performance of the various
\NDFS\ implementations in terms of their absolute timing and speedup
behavior. We will also compare them with the current
state-of-the-art algorithm in parallel symbolic model checking,
\OWCTYOTF~\cite{owcty-otf} by Barnat et al., which is a member
of the branch of \BFS-based algorithms.

Basically, it extends the {\em One-Way-Catch-Them-Young} algorithm
(\OWCTY~\cite{BrnoOWCTY}), with an initialization phase incorporated from
the {\em Maximal-Accepting-Predecessor} algorithm
(\MAP~\cite{BrnoMAP}). In a nutshell, \MAP\ iteratively
propagates unique node identifiers to successors. As soon as an
accepting state receives its own identifier, a cycle is detected.  \OWCTY\ is
based on topological sort and iteratively eliminates states that
cannot lie on an accepting cycle, because they have no predecessors.

These algorithms are generally based on \BFS, which is
more easy to parallelize than \DFS. However, these algorithms sacrifice
linear-time behavior and the on-the-fly property. The resulting
combination is linear-time for B\"uchi automata generated from the
class of weak LTL properties, and shows on-the-fly behavior for several cases.


\section{Experiments}\label{s:experiments}

We implemented multi-core Swarmed \NDFS\ and Alg.~\ref{alg:lndfs} and
Alg.~\ref{alg:endfs} in the 
multi-core backend of the \LTSmin\ model checking tool suite \cite{ltsmin,boosting,freelunch}.\footnote{Available on the \LTSmin\ website: \url{http://fmt.cs.utwente.nl/tools/ltsmin/}.}
We performed experiments on an AMD Opteron~8356 16-core 
($4 \times 4$ cores)
server with 64~GB RAM, running a patched Linux~2.6.32
kernel.
All tools were compiled using gcc~4.4.3 in 64-bit mode with high
compiler optimizations~(\texttt{-O3}).

We measured performance characteristics for 
all 453 models with properties of the \BEEM\ database \cite{beem} and 
compared the runs with
the best known parallel LTL model checking algorithm \OWCTYOTF\ as 
implemented in \Divine\ 2.5~\cite{barnat.brim.multicore,BBCR10}. In fact, we used the latest release available from the 
development repository on 23 March 2011, which was close to the 2.5 
version, except for a few relevant bug fixes.

Note that \OWCTYOTF\ has been implemented in \Divine, whereas all
\NDFS-based algorithms have been implemented in \LTSmin. This should
be taken into account when comparing absolute runtimes. 
\LTSmin\ implements a generic interface around the fast implementation
of the \post\ function of \Divine, resulting in sequential runtimes
that can be twice as slow. On the other hand, \LTSmin\ internally uses
shared hash tables, which are shown to scale better, at least for 
reachability~\cite{boosting}.

To account for the random nature of the algorithms, all experiments were
executed a total of 5 times. The data presented in the following
subsections reflect the average over those 5 experiments.

\subsection{EN{\small DFS} Benchmarks}
\begin{figure}[hpt!]
\begin{multicols}{2}
\hspace{-.1cm}
\includegraphics[width= 1.05\linewidth]{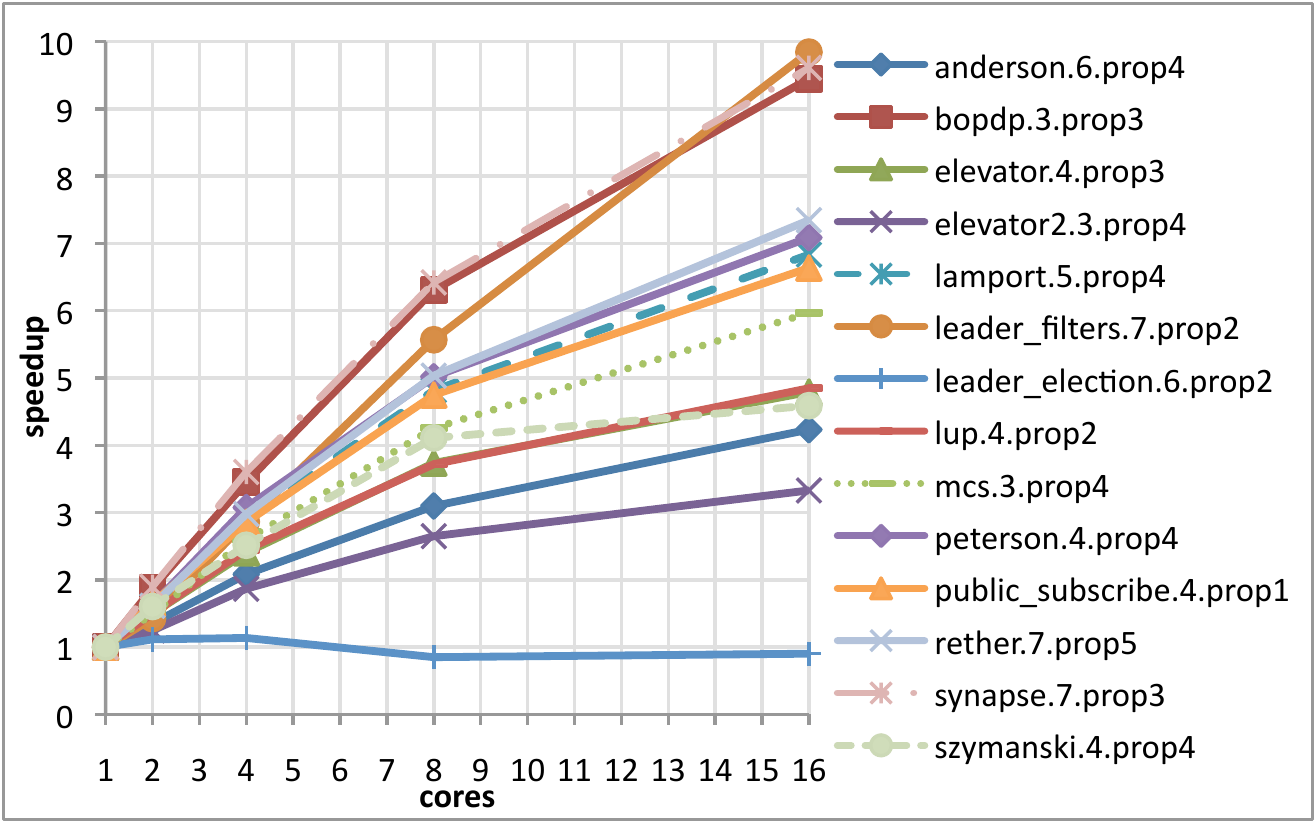}
\caption{Measured speedups for \ENDFS.}
\label{f:speedups_endfs}

\hspace{-.1cm}
\includegraphics[width= 1.05\linewidth]{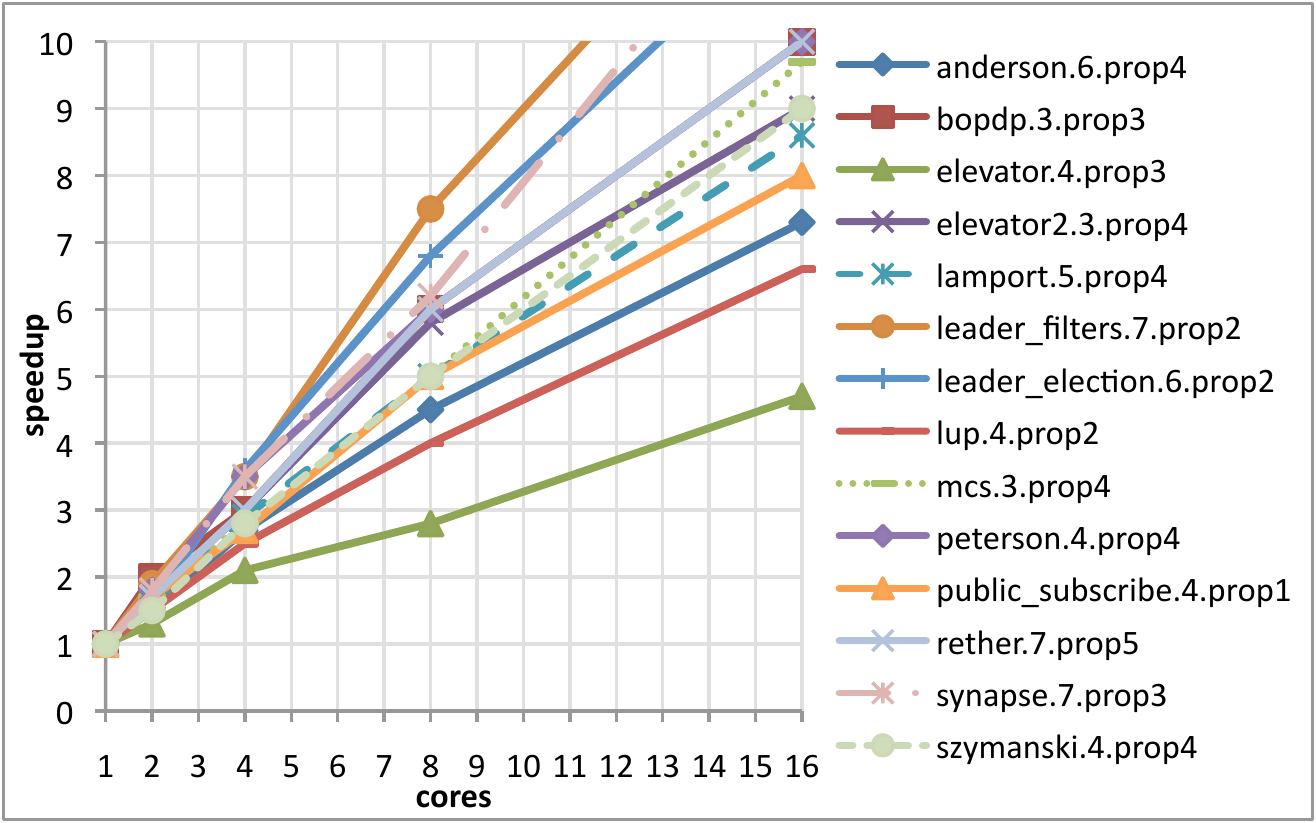}
\caption{Estimated speedups for \ENDFS\ from \cite{Evangelista11}.}
\label{f:speedups_paper}
\end{multicols}
\end{figure}

Evangelista et al. \cite{Evangelista11} used workload distribution
measurements to estimate the scalability of \ENDFS.
Fig.~\ref{f:speedups_paper} reflects their estimated speedups (the exact numbers were extracted from \cite{report}, which 
provides experiments for more models, but shows equal numbers to those 
reported in \cite{Evangelista11}).
Fig.~\ref{f:speedups_endfs} shows the speedups that we obtained by
measuring real runtimes of the algorithm.

A comparison with the estimated speedups shows that the trend of the lines
has been accurately predicted in most cases. 
A case by case comparison shows, however, that there is some divergence
between the exact numbers: models that scale well in ``synthetic''
benchmarks of Fig.~\ref{f:speedups_paper} as, for example,
\verb/anderson.6.prop4/, \verb/elevator2.3.prop4/, \verb/leader_election.6.prop2/ and \verb/szymanski.4.prop4/,
do not scale well in practice.
We have not investigated the source of these differences, but apparently the
amount of dangerous states is quite sensitive to implementation parameters.

\begin{figure}[b!]
\begin{multicols}{2}
\hspace{-.1cm}
\includegraphics[width= 1.05\linewidth]{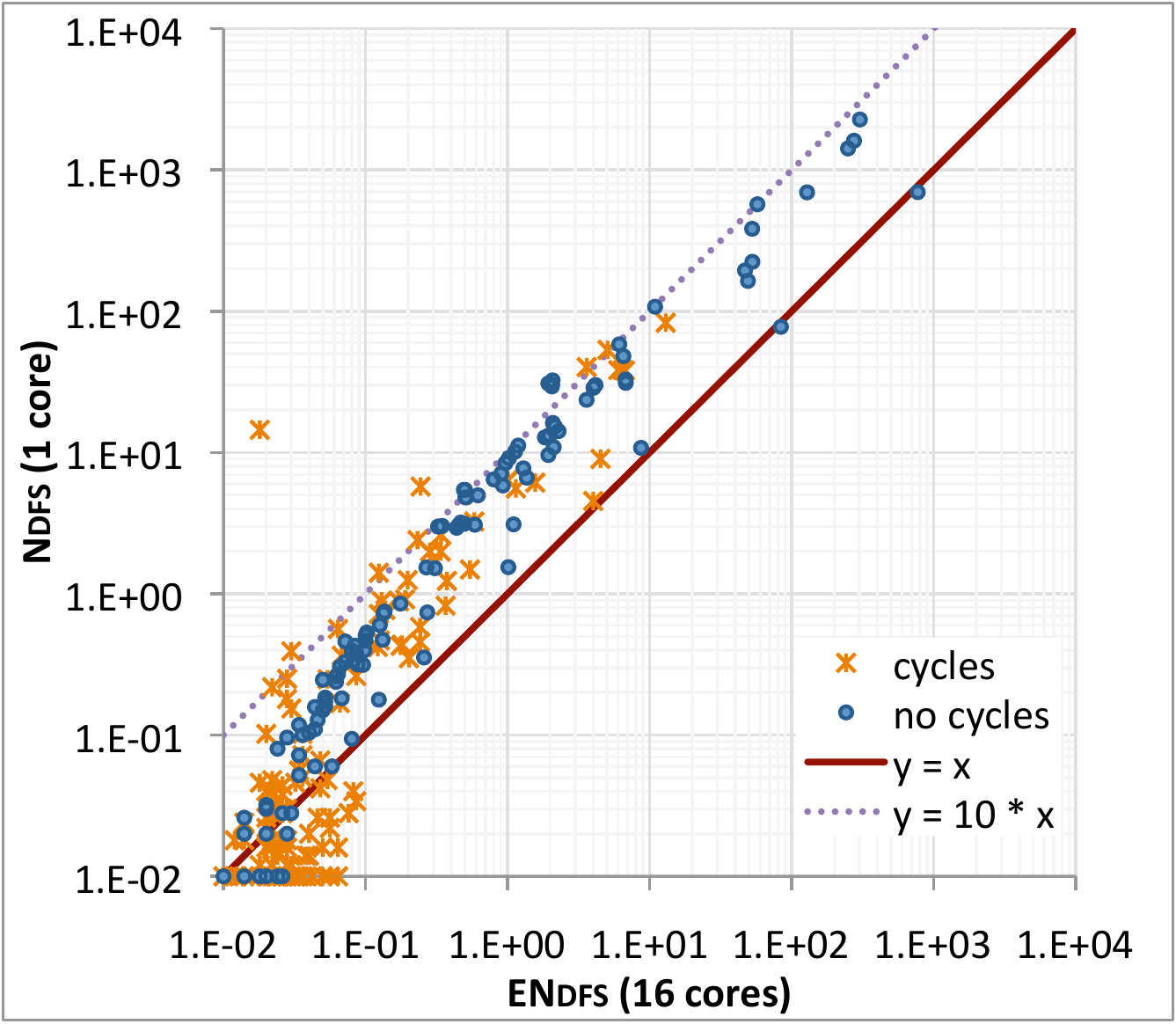}
\caption{\NDFS\ vs \ENDFS.}
\label{f:e_vs_n}
\hspace{-.1cm}
\includegraphics[width=1.05\linewidth]{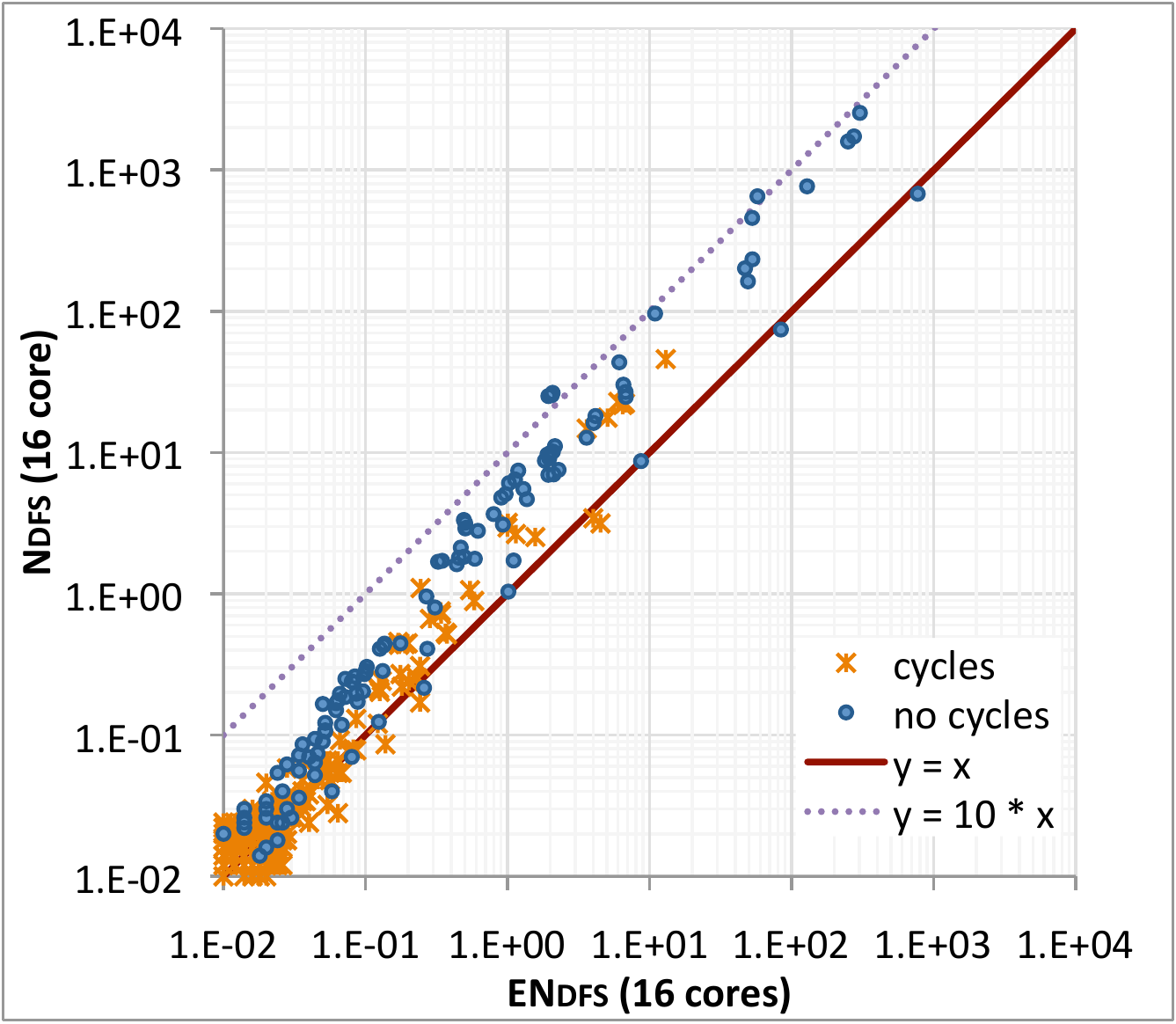}
\caption{Swarmed \NDFS\ vs \ENDFS.}
\label{f:e_vs_s}
\end{multicols}
\end{figure}

Fig.~\ref{f:e_vs_n} and Fig.~\ref{f:e_vs_s} compress the results from all 
models of the \BEEM\ database in log-log scatter plots. 
In both figures, we show models without accepting cycles as 
dots and models with these cycles as crosses.
Comparing \ENDFS\ to \NDFS\ in the first figure, 
we can distinguish good speedups  for the models with cycles,
while the other figure shows that \ENDFS\ even improves the results
of Swarmed \NDFS\ a little. In Section~\ref{s:discussion}, we investigate
and compare these effects more thoroughly, using a 
statistical reference model for random parallel search.
As for the models without accepting cycles, we see that most do scale
with \ENDFS, but hardly beyond a speedup of 10. 
Even though theoretically possible, we identified no cases
where the repair strategy of \ENDFS\ yields speed downs (in the worst
case, all workers can traverse the state space 4 times).

\begin{figure}[t!]
\begin{multicols}{2}
\hspace{-.1cm}
\includegraphics[width=1.05\linewidth]{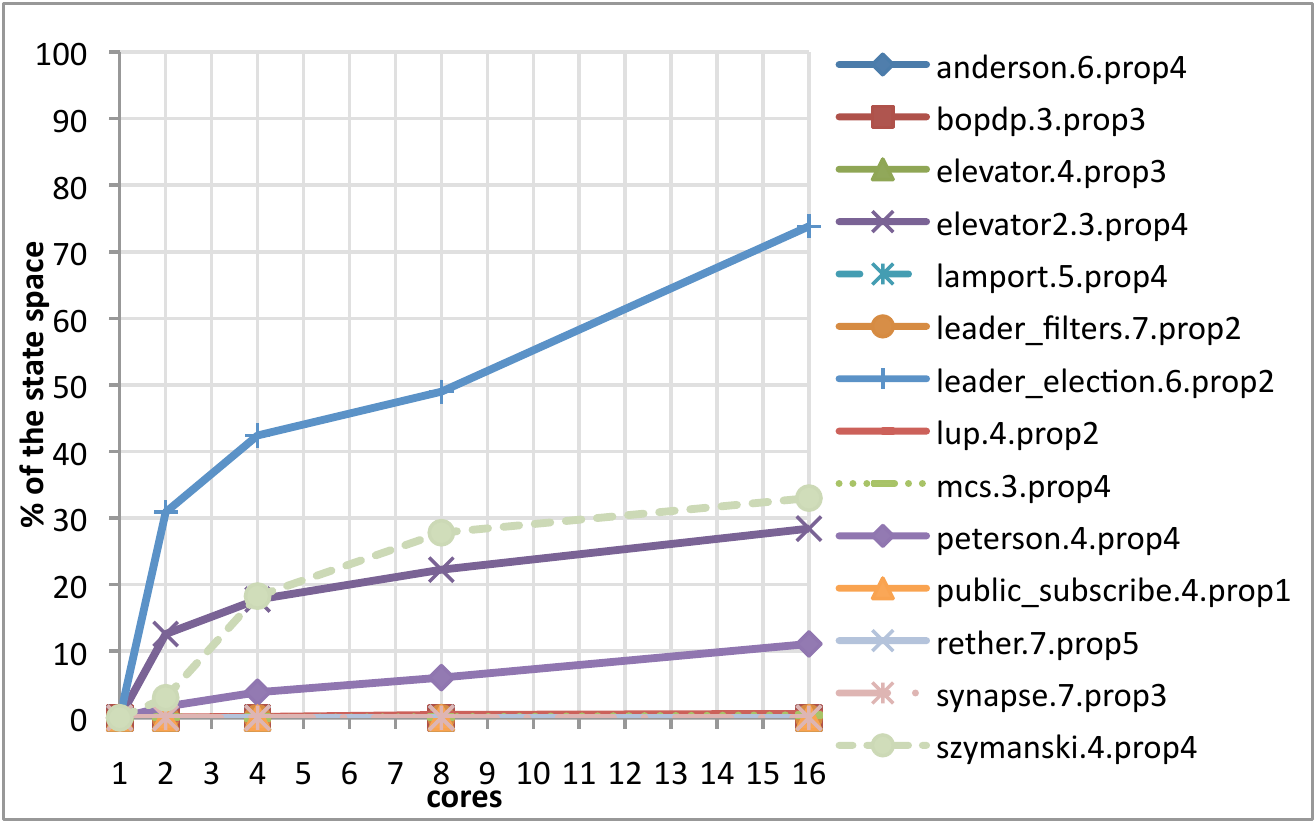}
\caption{State space coverage of \ENDFS\ repair.}
\label{f:repair}

\hspace{-.1cm}
\includegraphics[width=1.05\linewidth]{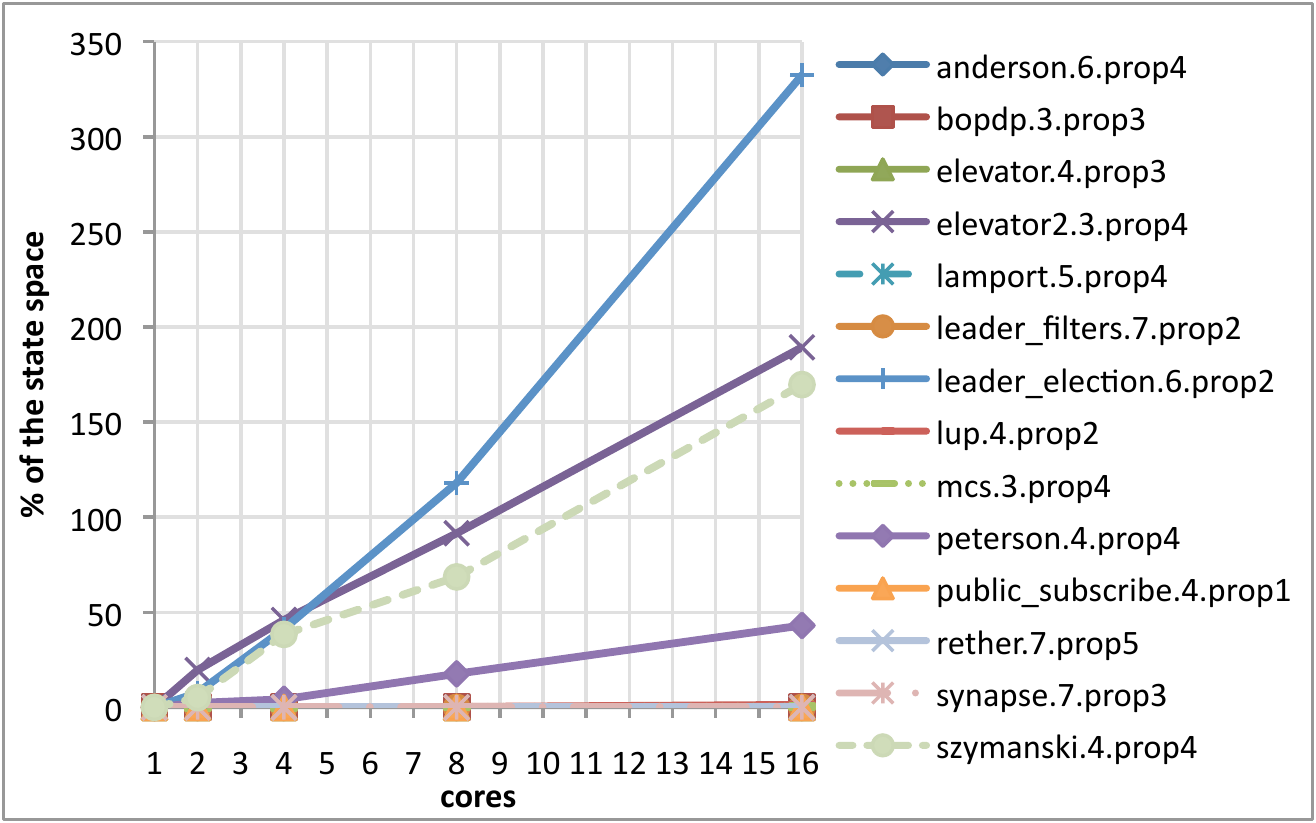}
\caption{Cumulative extra work due to repair.}
\label{f:repair2}
\end{multicols}
\end{figure}

We also investigated what caused some inputs to scale poorly.
Fig.~\ref{f:repair} shows the percentage of the state space that is 
covered by the repair procedure. As expected, a high percentage was
measured for all models with poor scalability.
Fig.~\ref{f:repair2} shows the cumulative additional work performed by all workers, 
by summing up the states visited by all workers in the repair procedure
and dividing by the total amount of states ($|\states|$).
It is worrisome that the need for repair can increase faster than the 
number of cores. This suggests that the \ENDFS\ may not scale to many-core 
systems.

\subsection{EN{\small DFS} versus LN{\small DFS}}

\begin{figure}[b!]
\begin{multicols}{2}
\hspace{-.1cm}
\includegraphics[width=1.05\linewidth]{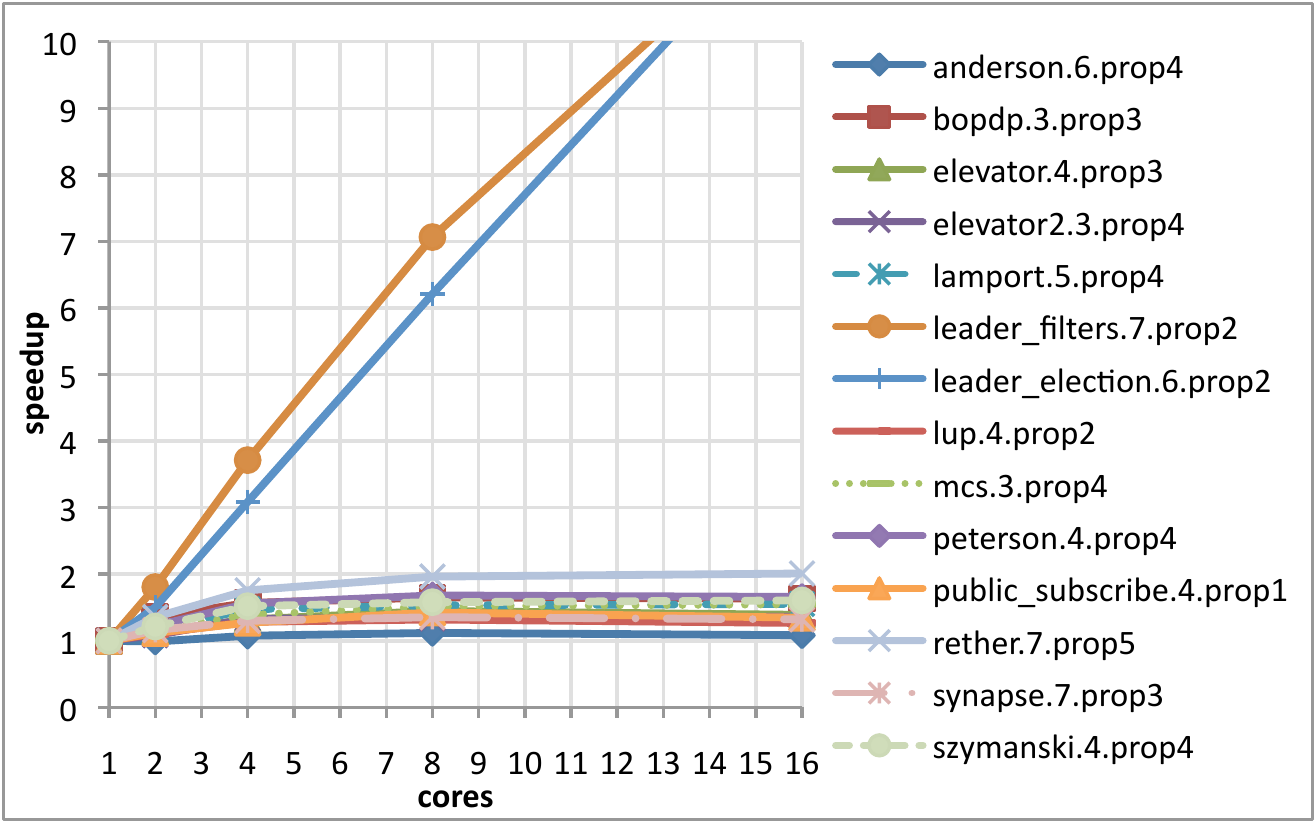}
\caption{Speedups \LNDFS.}
\label{f:speedups_lndfs}
\hspace{-.1cm}
\includegraphics[width=1.05\linewidth]{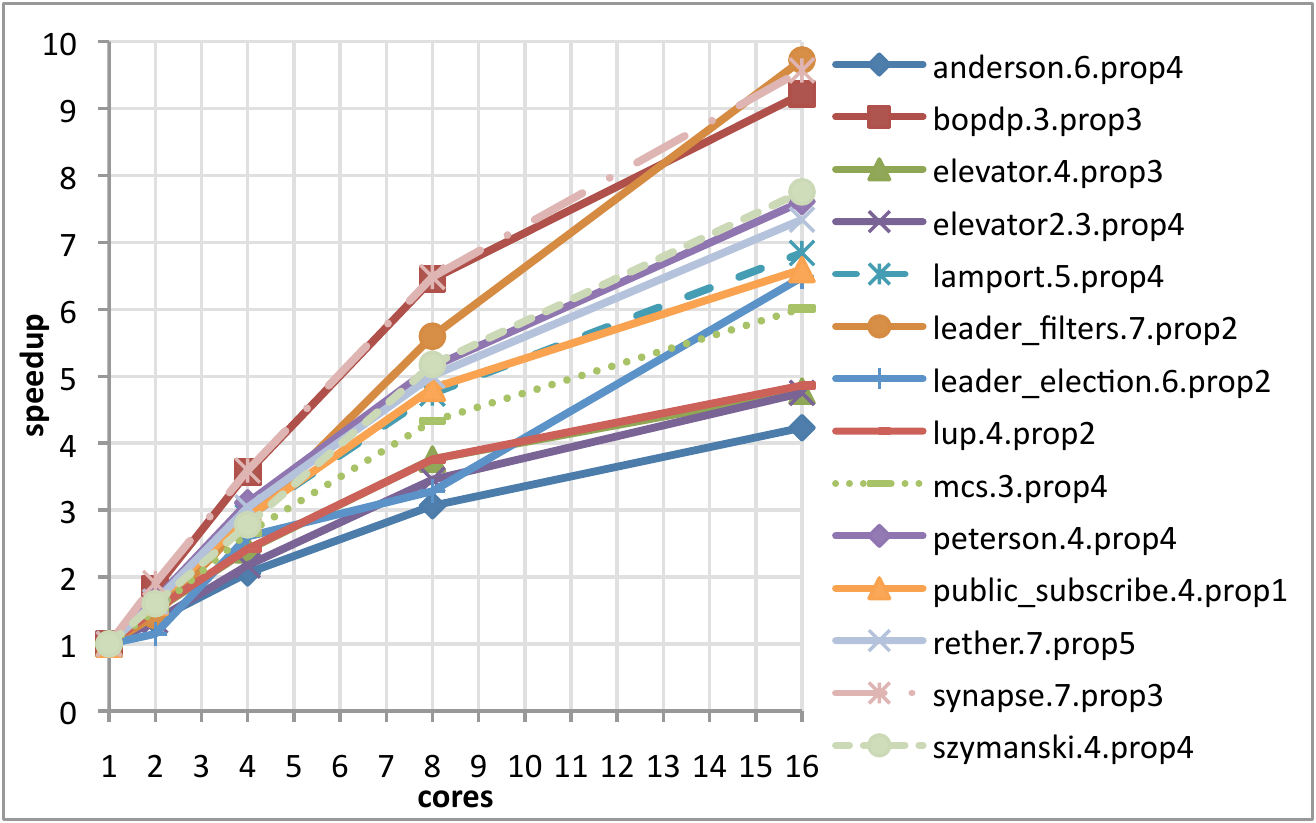}
\caption{Speedups \NMNDFS.}
\label{f:speedups_nmcdfs}
\end{multicols}
\end{figure}

\begin{figure}[t!]
\begin{multicols}{2}
\hspace{-.1cm}
\includegraphics[width= 1.05\linewidth]{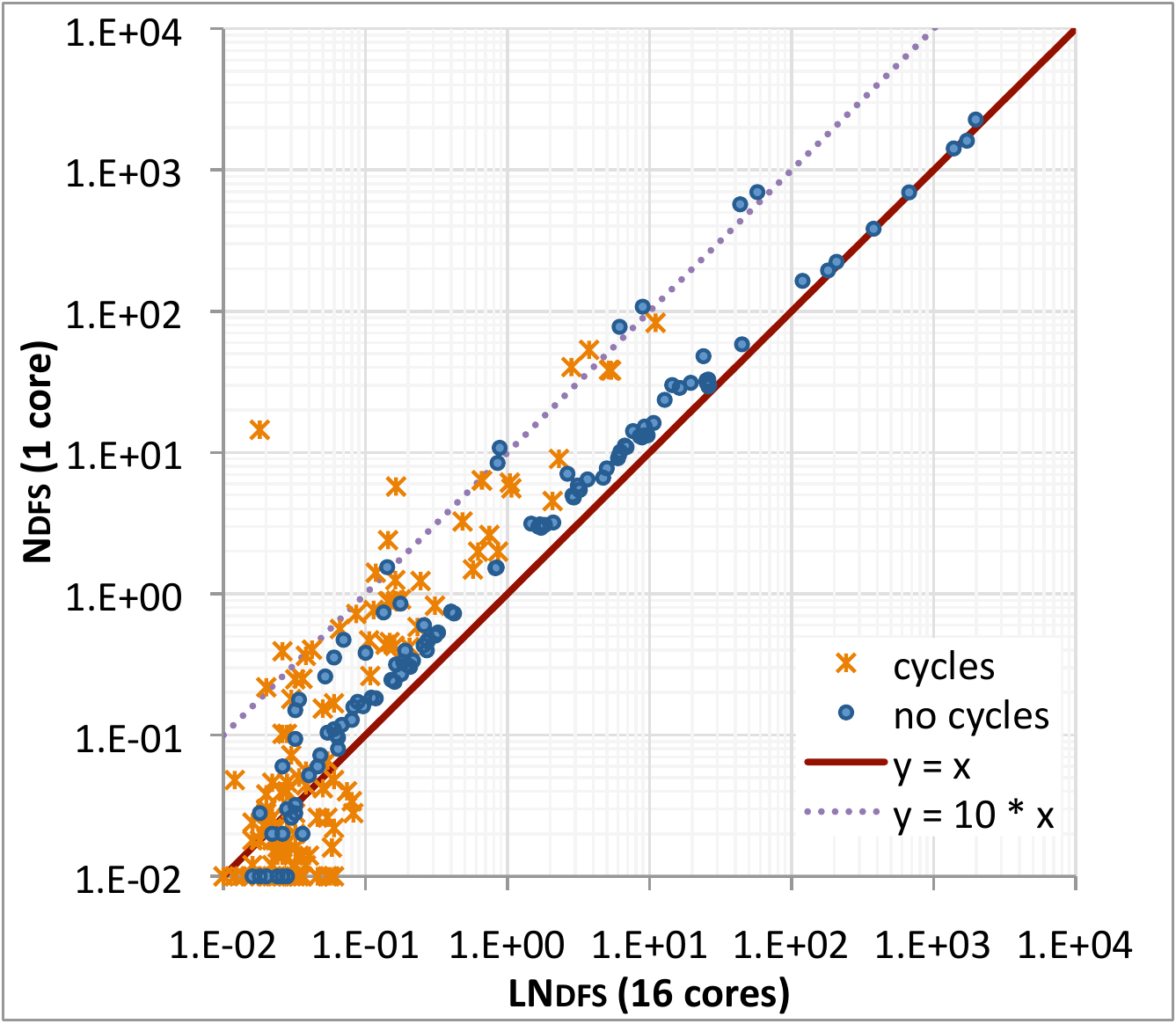}
\caption{\NDFS\ vs \LNDFS.}
\label{f:lndfs}
\hspace{-.1cm}
\includegraphics[width= 1.05\linewidth]{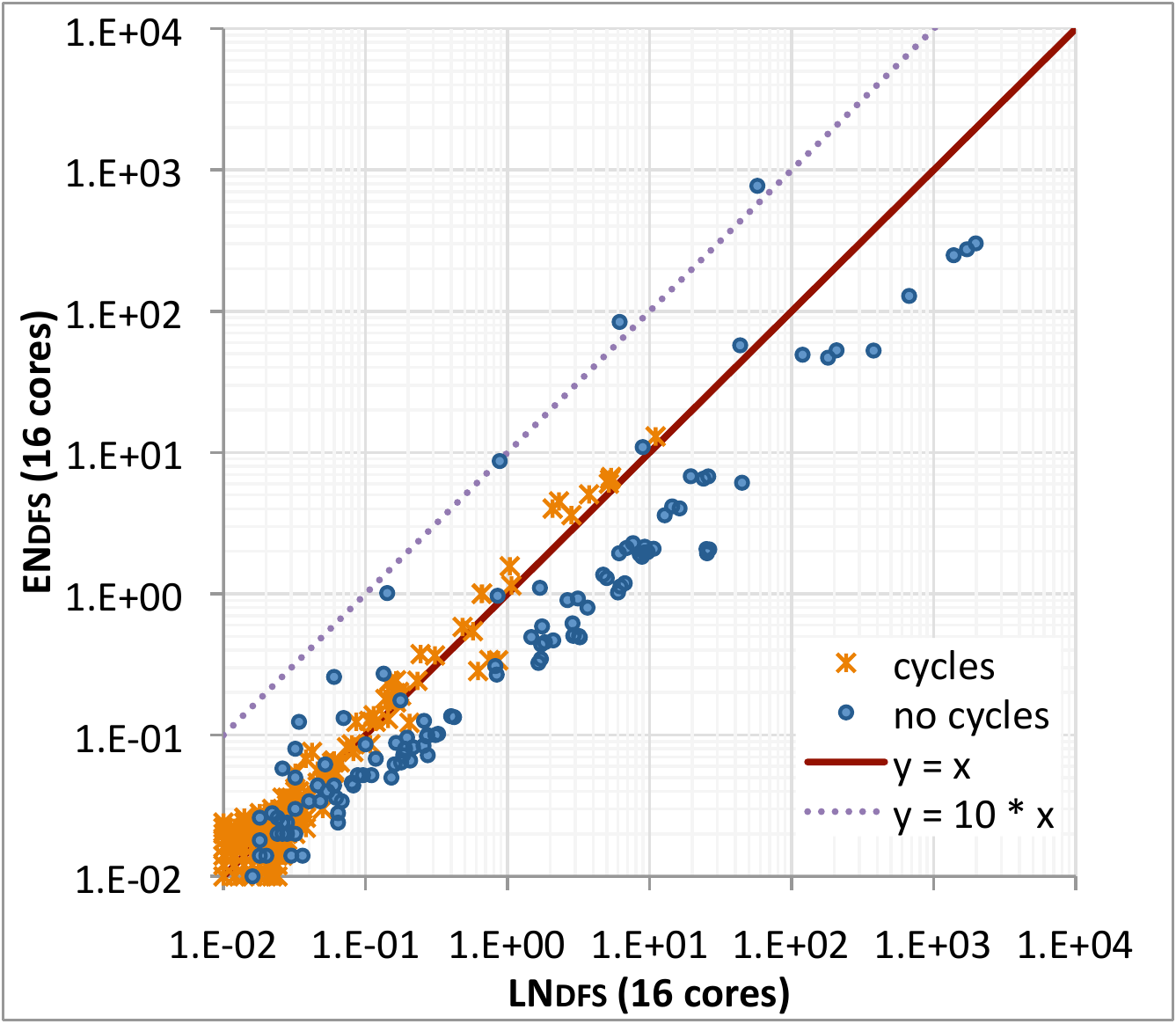}
\caption{\ENDFS\ vs \LNDFS.}
\label{f:e_vs_l}
\end{multicols}
\end{figure}

Fig.~\ref{f:speedups_lndfs} shows the speedups of the \LNDFS\ algorithm.
In this set of models, few scale well with this algorithm. The flat lines
represent models with relatively few states reachable from accepting 
states. In these cases, the algorithm can only color few states red,
thus limiting work sharing between the workers. As shown in \cite{Laarman11}, the
fraction of red states is indeed directly related to the speedup that
is obtained.
The two models \verb/leader_filters.7.prop2/ and
\verb/leader_election.6.prop2/ have state spaces that are colored entirely
red, and hence exhibit almost ideal linear speedups.
However, Fig.~\ref{f:lndfs} shows that only few models behave this 
ideally.
Unfortunately, in~\cite{Laarman11} we reported better speedups,
which we have now tracked down to an implementation error that led to 
too many red states.

When comparing \ENDFS\ to \LNDFS\ in Fig.~\ref{f:e_vs_l}, we witness 
a few ties (on the thick line), a few winners with \LNDFS\ and by far the most winners with 
\ENDFS. We looked up the models that draw a tie
and found that all of them scale with both algorithms. These are therefore
not in need of improvements.
Most interestingly, the models that scale well with
\LNDFS\ correspond to those that do not scale with \ENDFS. This indicates
that both algorithms are complementary. A fact that is indeed to be
expected, because the same accepting states that cause states to be 
colored red in \LNDFS, are potentially marked dangerous in \ENDFS. 
This motivated their combination as described in Section~\ref{s:nmndfs}.

\subsection{NM{\small C-NDFS} Benchmarks}

\begin{figure}[t!]
\begin{multicols}{2}
\hspace{-.1cm}
\includegraphics[width= 1.05\linewidth]{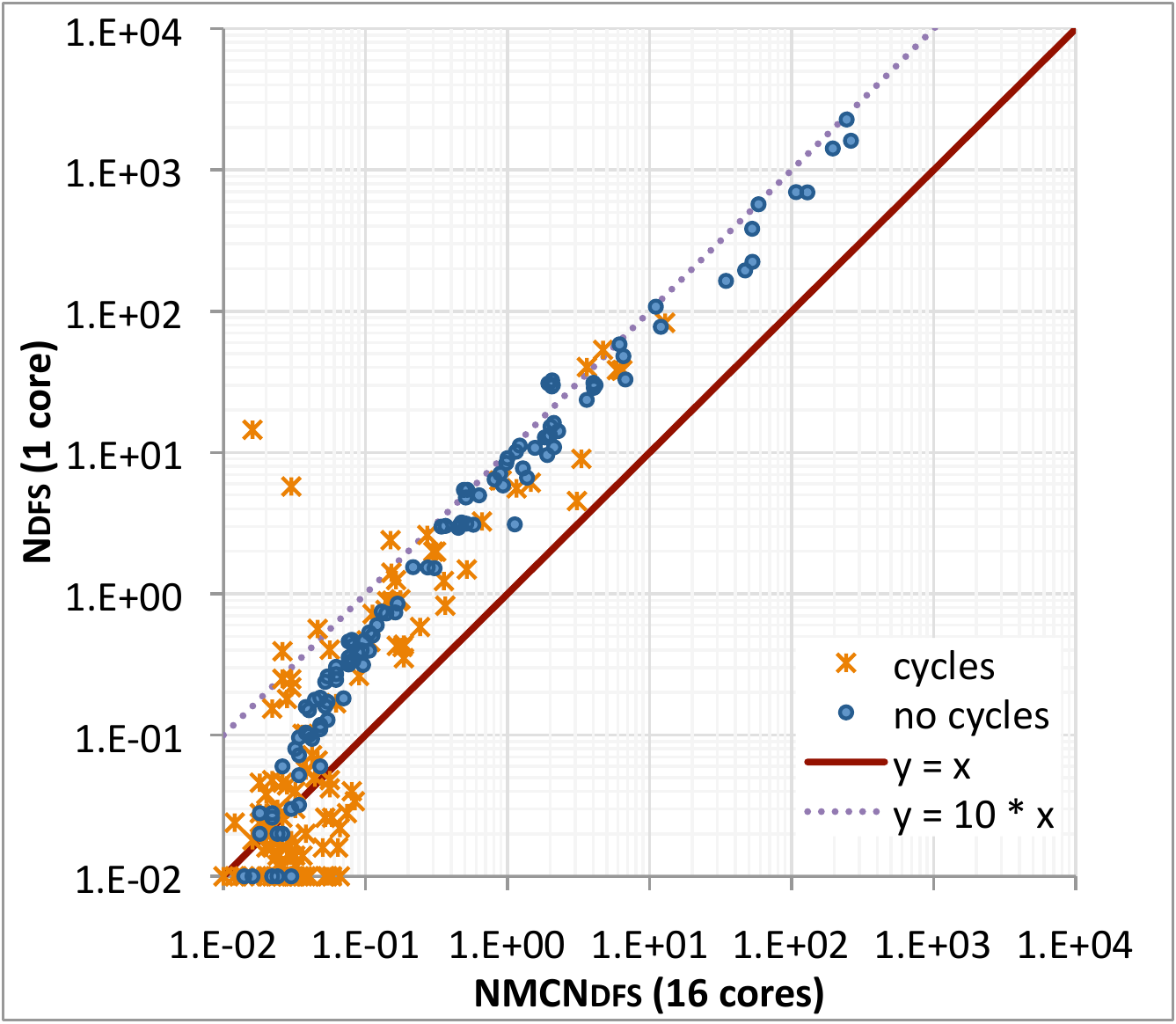}
\caption{\NDFS\ vs \NMNDFS.}
\label{f:speedup_nmcndfs}

\hspace{-.1cm}
\includegraphics[width=1.05\linewidth]{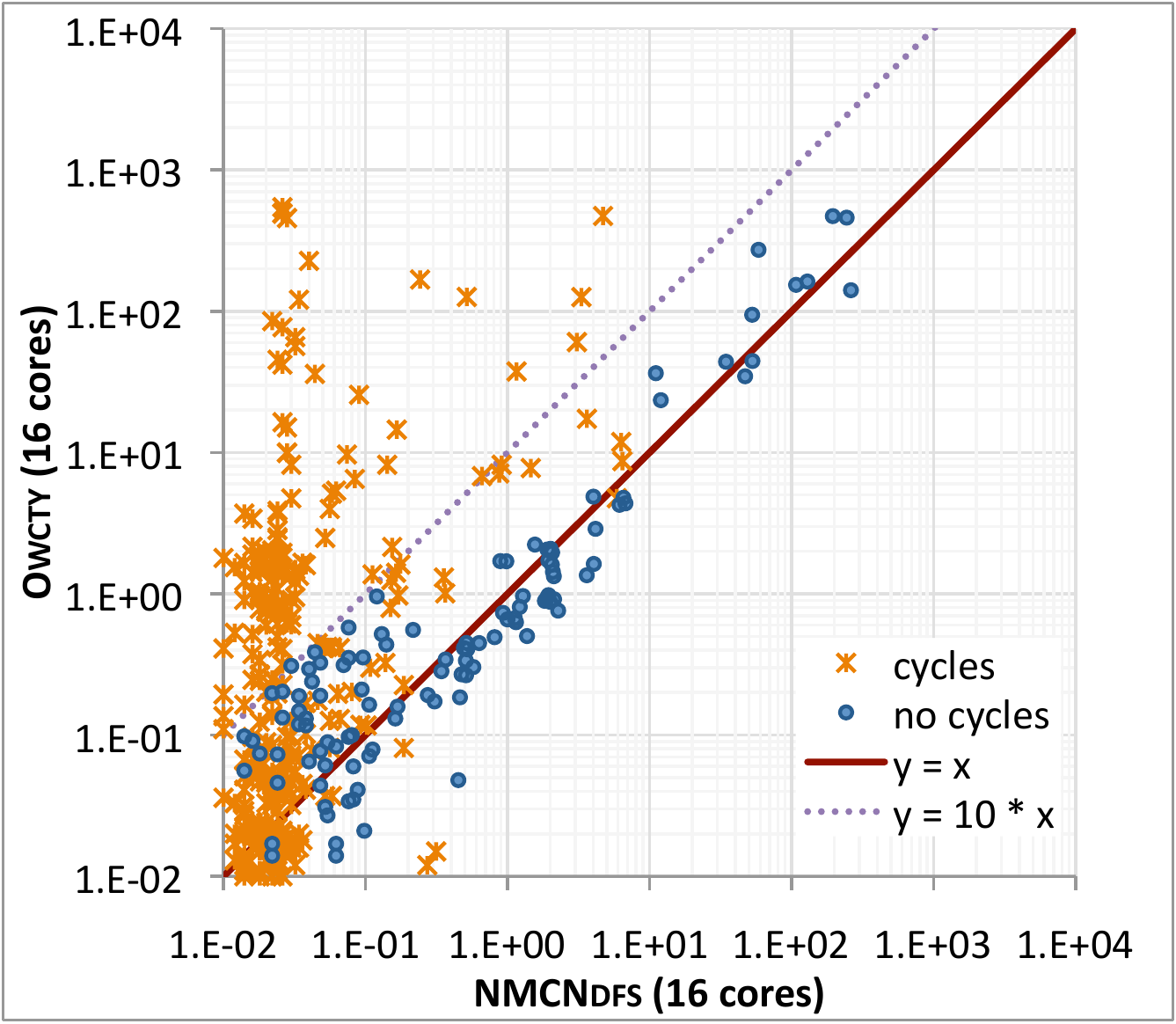}
\caption{\OWCTYOTF\ vs \NMNDFS.}
\label{f:n_vs_o}
\end{multicols}
\end{figure}

In this subsection, we investigate our proposal for the combination of
\ENDFS\ and \LNDFS\ into \NMNDFS.
Fig.~\ref{f:speedups_nmcdfs} shows that \NMNDFS\ improves upon the 
speedups of \ENDFS\ (see Fig.~\ref{f:speedups_endfs}), and
Fig.~\ref{f:speedup_nmcndfs} confirms that all models scale well
with the combined algorithm.

\begin{figure}[t!]
\begin{multicols}{2}
\hspace{-.1cm}
\includegraphics[width=1.05\linewidth]{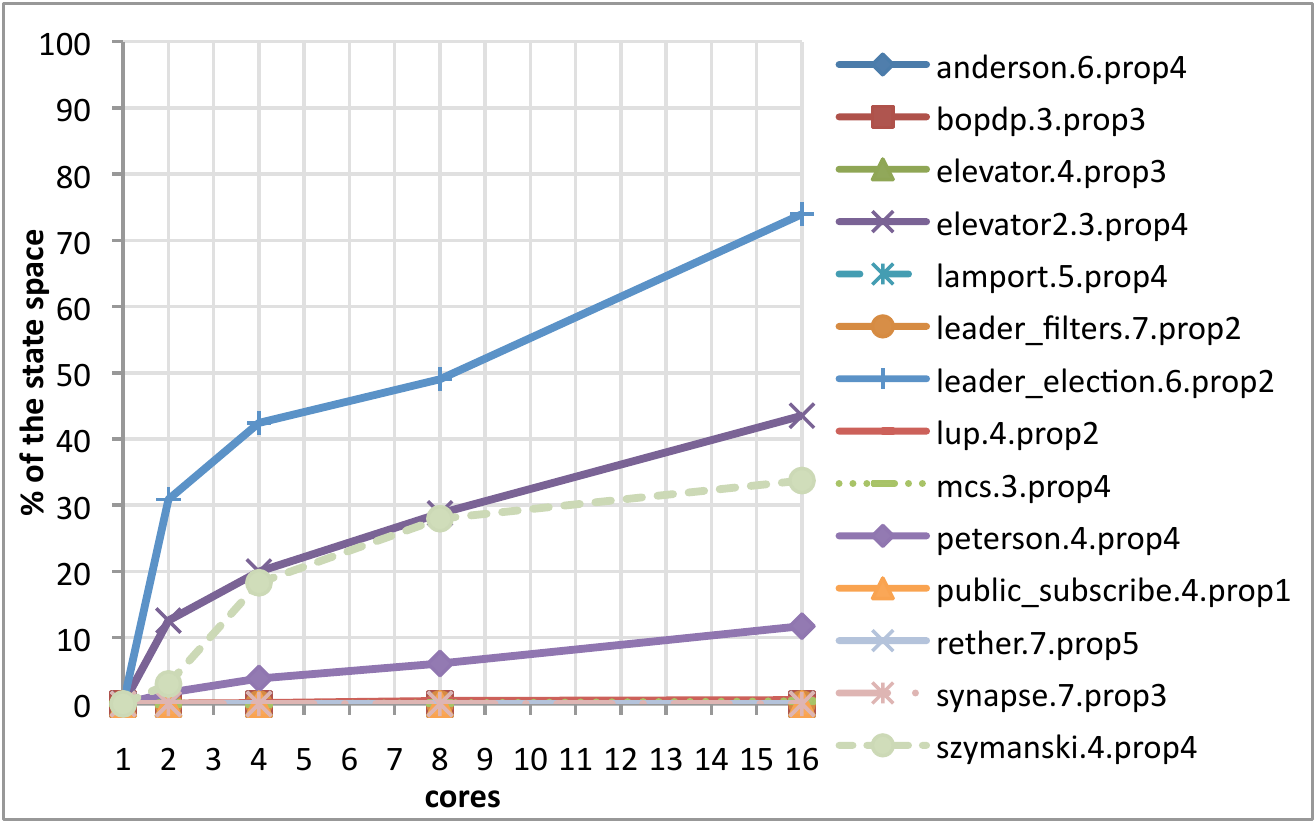}
\caption{Cumulative extra work due to \NMNDFS\ repair.}
\label{f:repair_l}
\hspace{-.1cm}
\includegraphics[width= 1.05\linewidth]{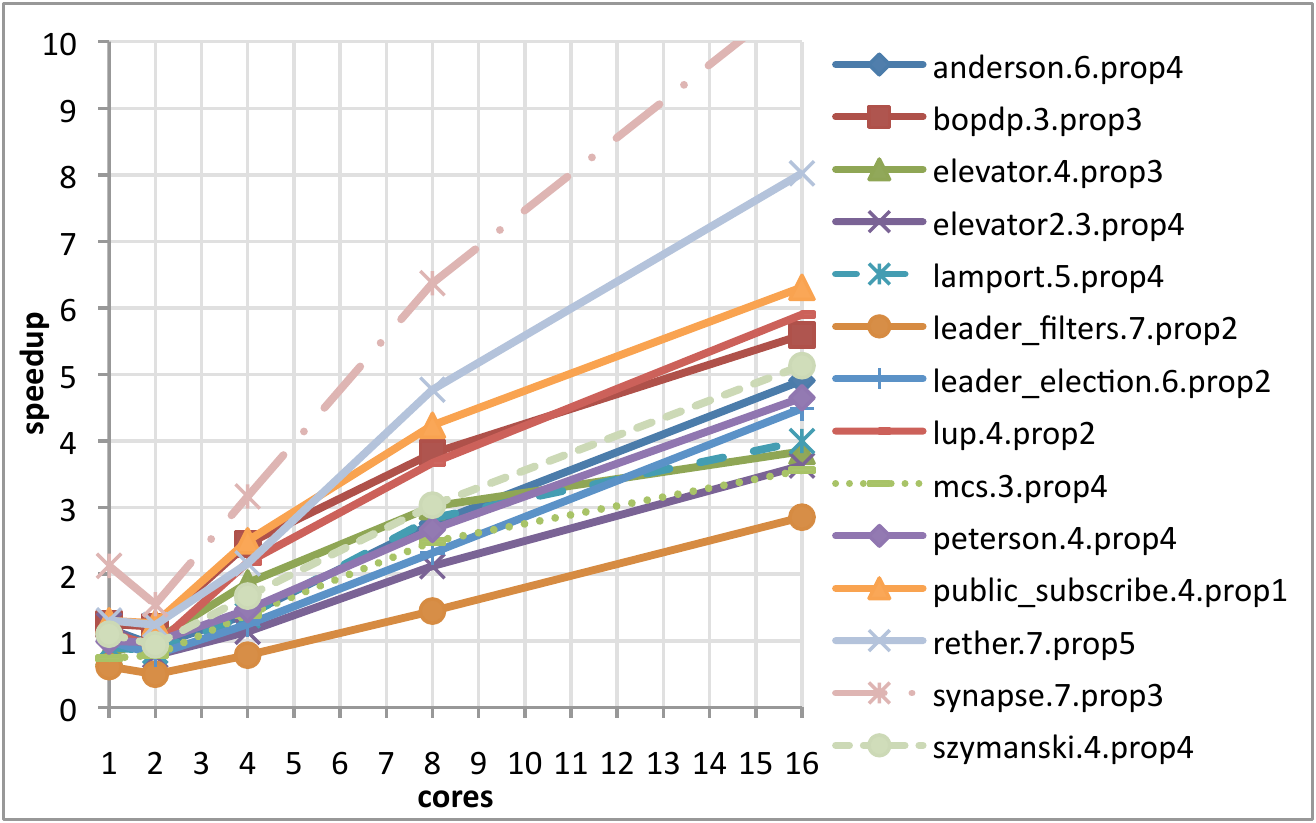}
\caption{Speedups \OWCTY.}
\label{f:speedups_owcty}
\end{multicols}
\end{figure}

For \NMNDFS, again, we also calculated the cumulative additional work as
a percentage of the state space in Fig.~\ref{f:repair_l}. The state space 
coverage by the repair procedure is almost equal to that of \ENDFS\ in
Fig.~\ref{f:repair}. We can then deduce that the repair work is 
parallelized well by \LNDFS, because the cumulative additional work is 
close to the percentage of state space coverage.
This can be explained by the fact that \LNDFS\ is always called on a
(dangerous) accepting state in \NMNDFS, which eventually leads to a red
coloring of the entire subgraph reachable from this accepting
state. Under these conditions \LNDFS\ can be expected to scale well.

\begin{figure}[hpt!]
\begin{multicols}{2}
\hspace{-.1cm}
\includegraphics[width= 1.05\linewidth]{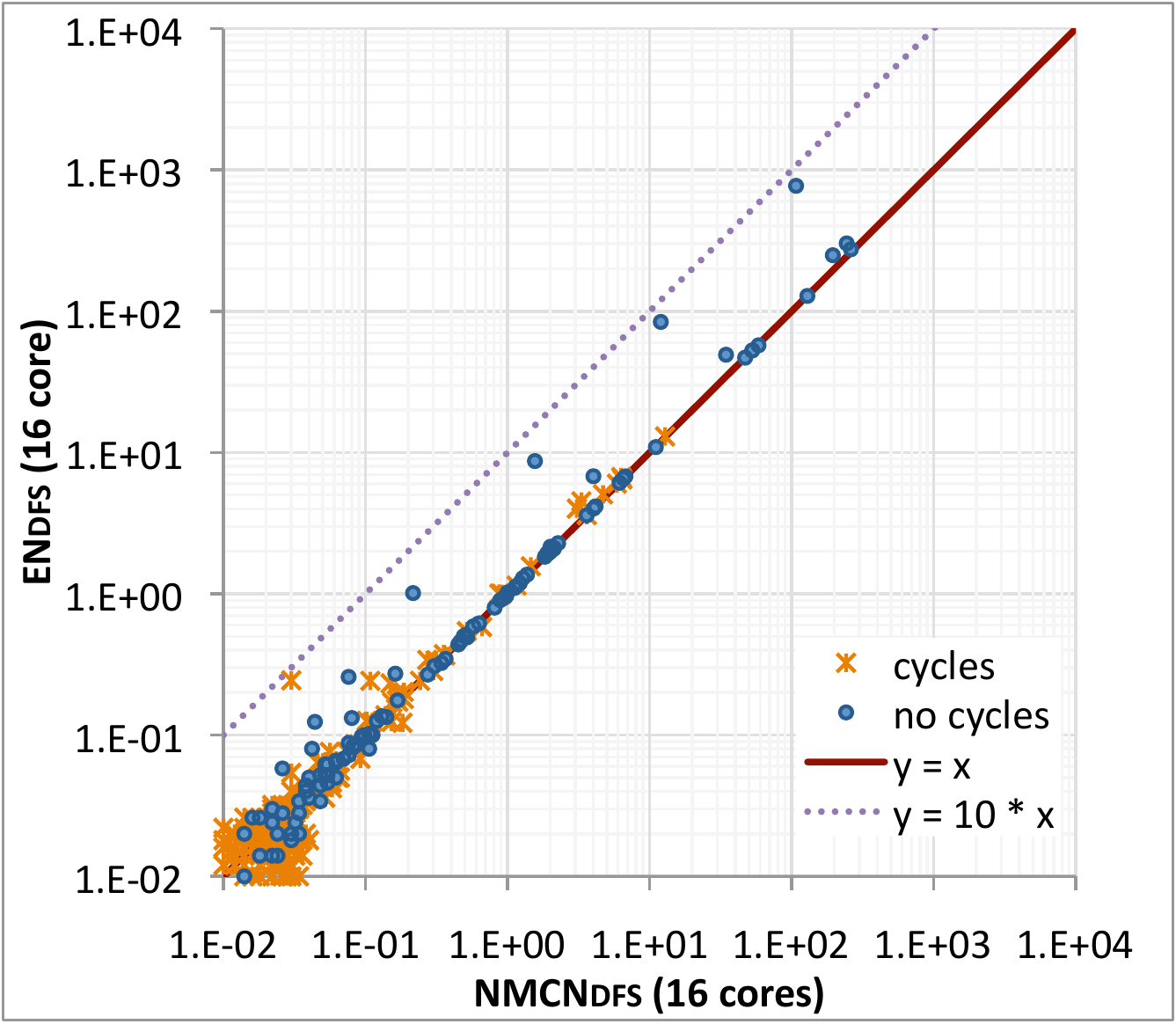}
\caption{\ENDFS\ vs \NMNDFS.}
\label{f:e_vs_nm}

\hspace{-.1cm}
\includegraphics[width=1.05\linewidth]{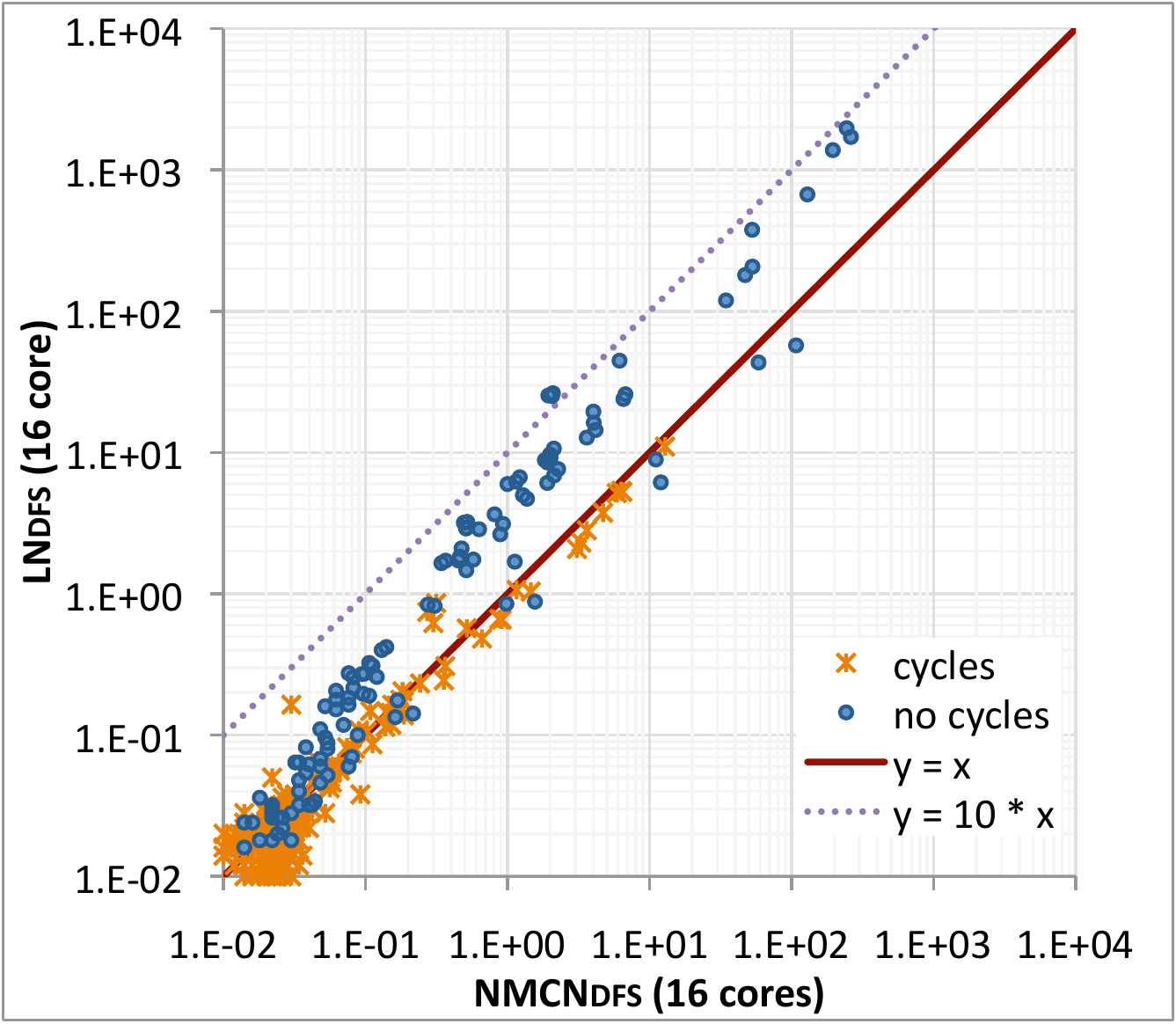}
\caption{\LNDFS\ vs \NMNDFS.}
\label{f:l_vs_n}
\end{multicols}
\end{figure}

We also checked whether the new combination
causes additional overhead, by comparing it directly with its predecessors in
Fig.~\ref{f:e_vs_nm} and Fig.~\ref{f:l_vs_n}. The first figure shows that
no model runs faster with \ENDFS\ than with \NMNDFS, although in a few
examples \LNDFS\ wins, as can be seen in the latter figure. This confirms
that \LNDFS\ and \ENDFS\ are complementary and their combination represents
the best from both worlds. Indeed, the combination ensures that
for all inputs some speedup is obtained.

\newpage
\subsection{Parallel N{\small DFS} versus O{\small WCTY}-M{\small AP}}

\begin{figure}[hpt!]
\begin{multicols}{2}
\hspace{-.1cm}
\includegraphics[width= 1.05\linewidth]{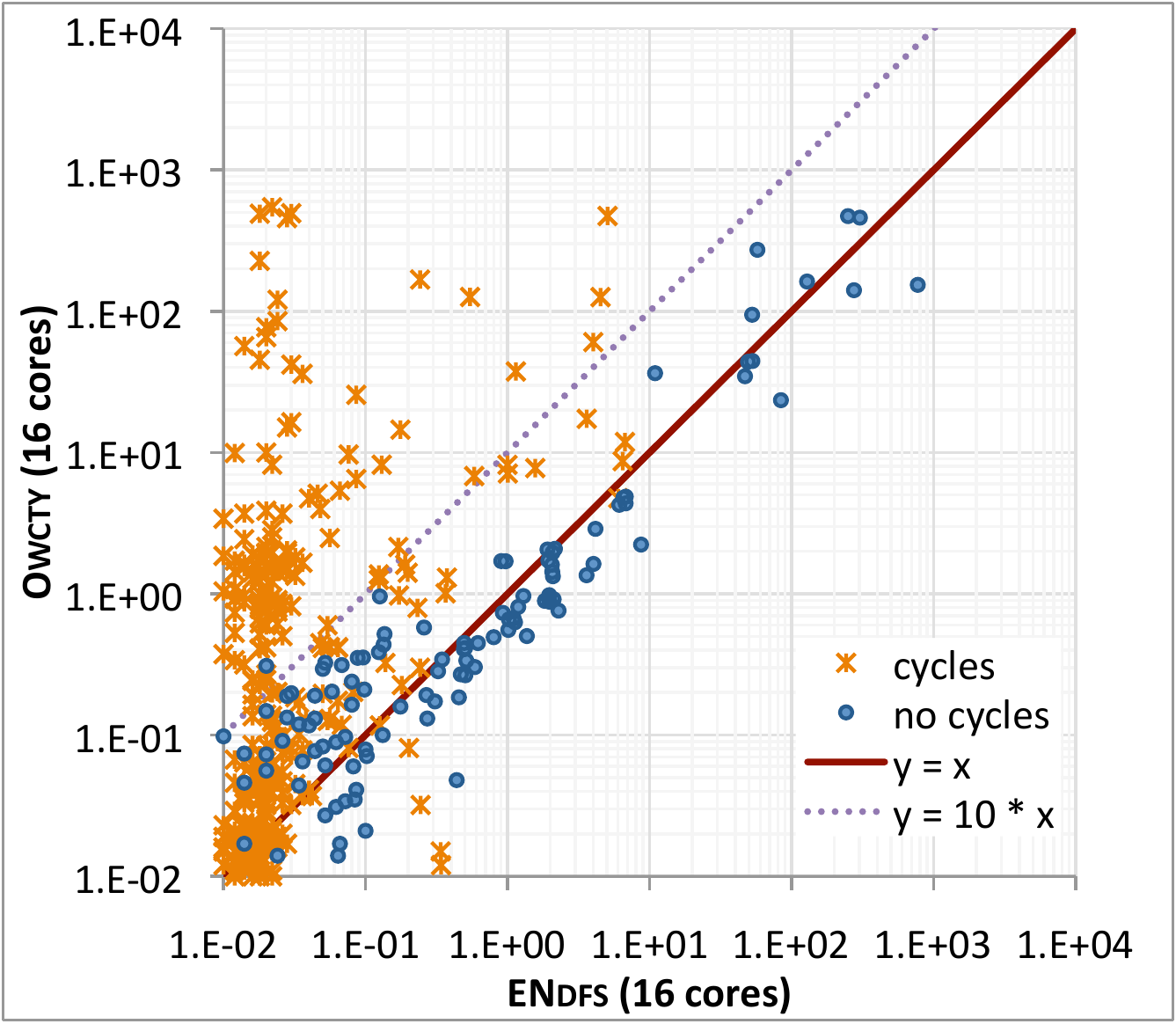}
\caption{\OWCTYOTF\ vs \ENDFS.}
\label{f:endfs2}

\hspace{-.1cm}
\includegraphics[width= 1.05\linewidth]{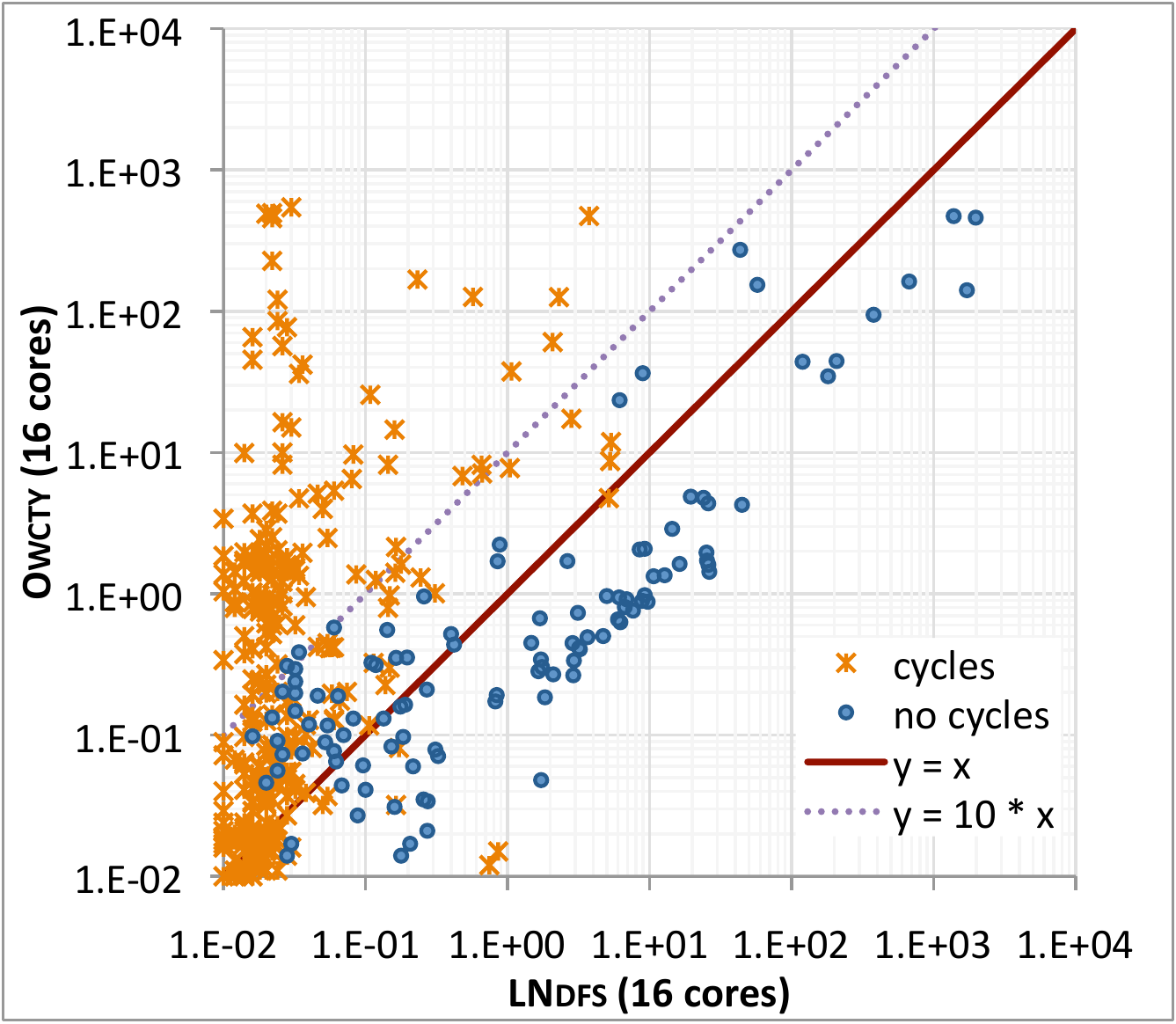}
\caption{\OWCTYOTF\ vs \LNDFS.}
\label{f:lndfs2}
\end{multicols}
\end{figure}

Fig.~\ref{f:n_vs_o} compares \NMNDFS\ with \OWCTYOTF.
The comparison figures show
that the heuristic \hbox{on-the-fly} method of \OWCTYOTF\
is no match for the truly on-the-fly parallel \NDFS\ algorithms. As for
the models without accepting cycles, we can conclude that currently
\NMNDFS\ provides a good match for \OWCTYOTF, in particular for the larger models.
For the sake of completeness, we present here 
Fig.~\ref{f:endfs2},~\ref{f:lndfs2}, which show a comparison between
\ENDFS/\LNDFS\ and \OWCTYOTF. Furthermore,
Fig.~\ref{f:speedups_owcty} shows the absolute speedups of \OWCTYOTF\
using the sequential \NDFS\ runtimes as the base~case.

\newpage

\newpage


\section{Discussion on Parallel Random Search}\label{s:discussion}

As explained in Section~\ref{s:algorithms}, the multi-core
\NDFS\ algorithms use a randomized \post\ function to direct workers
to different regions of the state space. In this section, we want to
explain the speedup for models with accepting cycles. In particular,
we want to distinguish the effect of parallel random search, from
the effect of the clever work sharing algorithms.

Our starting point is a simple statistical model as found
in~\cite{HyvarinenJN08}. We view $\NDFS(\buchi,X)$ as an algorithm
that runs on B\"uchi automaton $\buchi$ with random seed $X$,
influencing the order of traversing successors. We ran
$\NDFS(\buchi,X)$ 500 times with random $X$ on a number of B\"uchi
automata $\buchi$. Each time, we measured $f(\buchi,X)$, the time that it takes
for $\NDFS(\buchi,X)$ to detect an accepting cycle.

\begin{figure}[b]
\begin{center}
\includegraphics[width=0.77\linewidth]{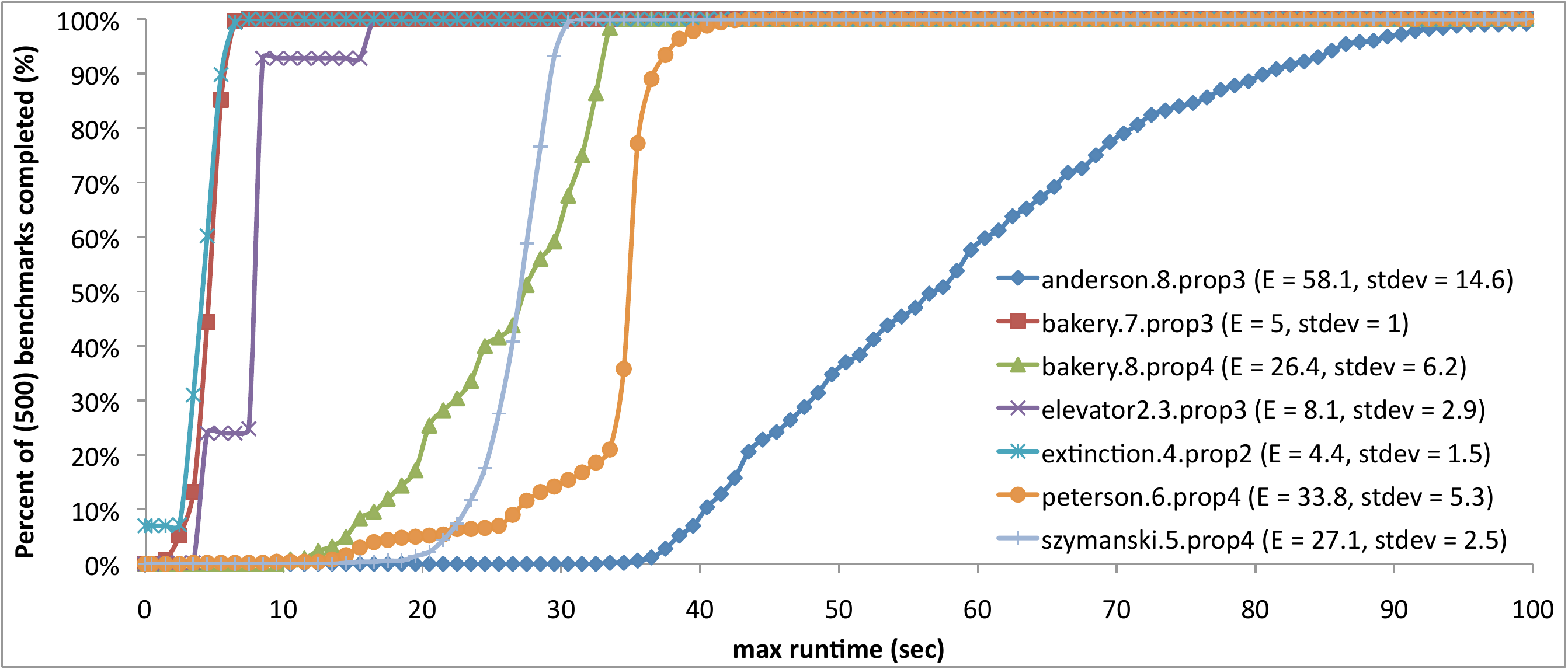}
\caption{Cumulative probability distribution of finding a bug (measured for 1 worker).}
\label{f:histo1}
\end{center}
\medskip
%
\begin{center}
\includegraphics[width=0.77\linewidth]{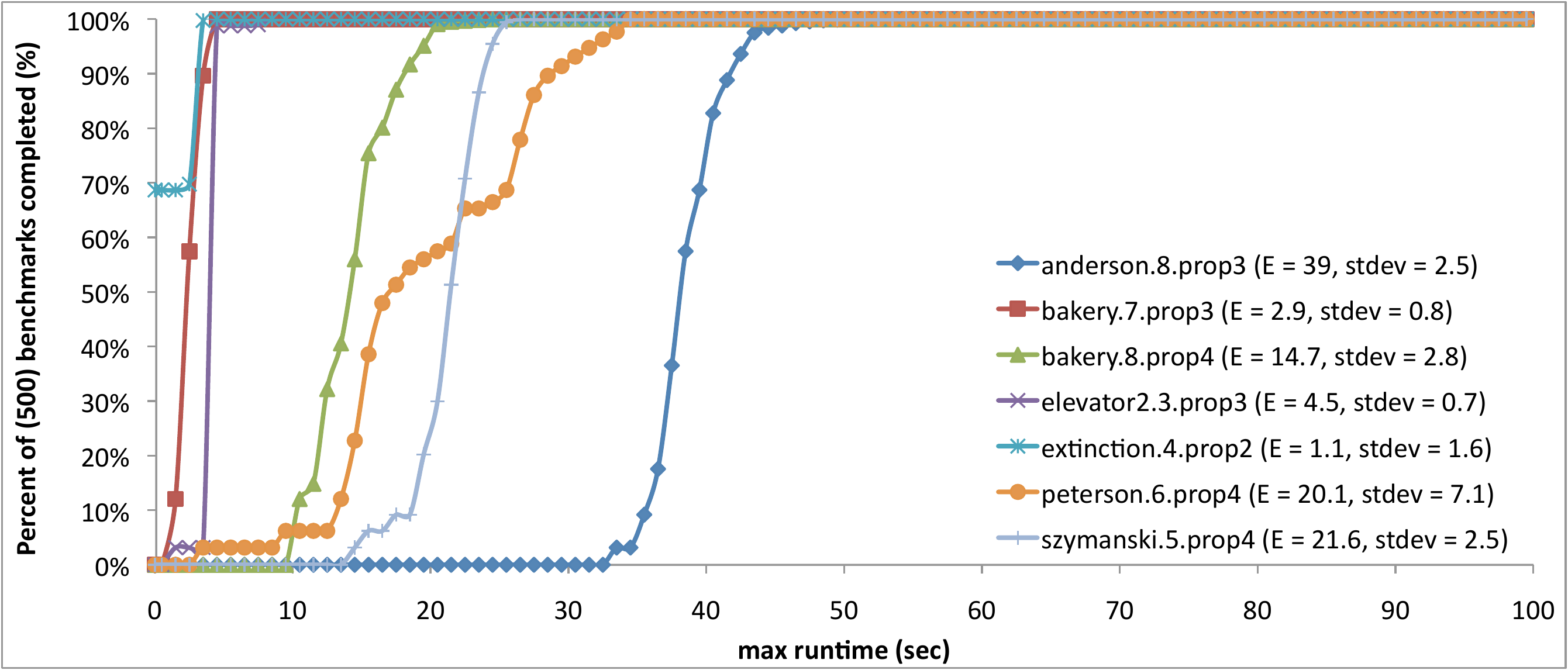}
\caption{Cumulative probability distribution of finding a bug (calculated for 16 workers).}
\label{f:histo2}
\end{center}
\end{figure}

In Figure~\ref{f:histo1}, we show the cumulative probability $F(\buchi,t)$ 
that one \NDFS\ worker will detect an accepting cycle in less than $t$
seconds for some examples from the \BEEM\ database.
We can also define $F_N(\buchi,t)$ as the cumulative probability that a swarm of $N$
independent workers will find an accepting cycle within $t$ seconds.
Figure~\ref{f:histo2} shows $F_{16}(\buchi,t)$ for the same automata. 
We also computed the expected time to completion and the standard deviation.
The new distribution can be easily computed as:
\[ F_N(\buchi,t) = 1 - (1 - F(\buchi,t))^N \]

From Fig.~\ref{f:histo1} and Fig.~\ref{f:histo2}, we observe that considerable 
gains can be expected from a simple parallelization as in Swarmed \NDFS.
It also shows that the actual speedup depends highly on the models:
when all runs find an accepting cycle in about the same time
(indicated by plateaus connected by a steep curve), the expected gain
is much less than when the curve is flatter, as is the case for
{\tt anderson.8.prop3}, {\tt bakery.8.prop4} and {\tt peterson.6.prop4}.

\begin{table}[pb!]
\caption{Runtimes and speedups of bug hunting using embarrassingly 
parallel (randomized) \NDFS\ and \LNDFS.
The first two columns of the table present the expected completion time derived from 500 
sequential experiments for 1 and 16 cores.
The other columns give parallel runtimes for, respectively, a distributed 
implementation, our randomized shared-memory
implementation~\cite{Laarman11},
and another shared-memory implementation 
using the fresh successor heuristic. The second row gives~the~speedups.}
\begin{center}
\begin{tabular}{|p{2.2cm}|l|r|r|r|r|r|r|r|}

\cline{3-9}
\multicolumn{2}{c}{} &
\multicolumn{5}{|c}{\NDFS} &
\multicolumn{2}{|c|}{\LNDFS} \\

\cline{3-9}
\multicolumn{2}{l}{} 	&
\multicolumn{1}{|p{5ex}}{1~core}		&
\multicolumn{4}{|c}{16 core}		&
\multicolumn{2}{|c|}{16 core} 	\\

\cline{2-9}
\multicolumn{1}{c}{} &
\multicolumn{1}{|c}{model} 	&
\multicolumn{1}{|c}{\begin{sideways}Statistical model~ \end{sideways}}&
\multicolumn{1}{|c}{\begin{sideways}Statistical model~ \end{sideways}}&
\multicolumn{1}{|c}{\begin{sideways}Distributed\end{sideways}}	&
\multicolumn{1}{|c}{\begin{sideways}Shared Memory\ \end{sideways}}&
\multicolumn{1}{|c}{\begin{sideways}Heuristic\end{sideways}}	&
\multicolumn{1}{|c}{\begin{sideways}Shared Memory\ \end{sideways}}&
\multicolumn{1}{|c|}{\begin{sideways}Heuristic\end{sideways}} 	\\

\hline
\multirow{7}{3cm}{Runtimes (sec)} 
&anderson.8.prop3	&	58.1&	39.0& 39.3	& 39.4	& 9.6	& 8.6	& 3.1\\
&bakery.7.prop3		&	5.0	&	2.9	& 2.9	& 2.1	& 0.6	& 0.8	& 0.3\\
&bakery.8.prop4		&	26.4&	14.7& 13.6	& 12.9	& 0.6	& 1.9	& 1.1\\
&elevator2.3.prop3	&	8.1	&	4.5	& 4.2	& 2.6	& 0.7	& 2.1	& 0.2\\
&extinction.4.prop2	&	4.4	&	1.1	& 0.8	& 0.5	& 0.0	&0.0	& 0.0\\
&peterson.6.prop4	&	33.8&	20.1& 24.2	& 16.7	&12.5	&2.5	& 2.2\\
&szymanski.5.prop4	&	27.1&	21.6& 20.9	& 19.4	& 0.0	&3.3	& 0.0\\
\hline
\multirow{7}{3cm}{Speedups}
&anderson.8.prop3	&	&	1.5	&	1.5 & 1.5	& 6.1	& 6.7	& 18.5	\\
&bakery.7.prop3		&	&	1.7 &	1.7 & 2.4	& 8.6	& 6.3	& 15.2	\\
&bakery.8.prop4		&	&	1.8	&	1.9 & 2.0	& 45.7	& 14.1 	& 23.2	\\
&elevator2.3.prop3	&	&	1.8	&	1.9 & 3.1	& 11.7	& 3.8	& 41.8	\\
&extinction.4.prop2	&	&	4.1	&	5.9 & 8.9	& ??	& ??	& ??	\\
&peterson.6.prop4	&	&	1.7	&	1.4 & 2.0	& 2.7	& 13.5	& 15.6	\\
&szymanski.5.prop4	&	&	1.3	&	1.3 & 1.4	& ??	& 8.3	& ??	\\
\hline
\end{tabular}
\end{center}
\label{t:speedups}
\end{table}

\medskip
Next, we want to compare our actual implementation with these predictions.
To this end, we compared the expected completion times with actual
completion times, averaged over 5 runs. We collected this information in
Table~\ref{t:speedups}. In the first two columns (Statistical model), we
copied the averages from Fig.~\ref{f:histo1},~\ref{f:histo2} for 1 and 16
workers, and computed
the expected speedup. Note that this speedup for 16 workers is way below 16.
Next, we experimented with four different scenarios described below.

The next column (Distributed), corresponds to Swarmed \NDFS\ as it would
run on different machines in a GRID.
Here the only synchronization would
be to terminate all workers as soon as the first worker has detected a
cycle. The runtimes denote the completion time for the 
earliest run out of 16 independent workers; we again provide the average
from 5 experiments. The corresponding speedups match closely to the
predicted ones from the statistical model.

Next, we ran the experiments on the multi-core machine with 16 cores
described before. Now the workers share the basic infrastructure.
This is the same setting as the multi-core Swarmed \NDFS\ from the 
previous section. For
instance, all states will be stored only once in a shared hash table.
Also, several workers now share information in the L2 cache. On the
other hand, they might now suffer from cache coherence overhead or
memory bus contention. The figures under ``Shared Memory'' show that the 
speedups in a multi-core environment are slightly better than on independent
machines (Distributed). 

On multi-core machines it becomes easier to share information, in
order to guide different workers into different parts of the state
space. In that case, one would expect better speedup figures.
We did an experiment with what we call the {\em fresh
successor heuristic}. Here a worker will randomly select
a globally unvisited successor if that exists, otherwise it randomly 
selects any successor. As the column Heuristic shows, this can
dramatically improve the speedup of 16 workers. In some cases, each
time an accepting cycle was found in such a small instant that a 
meaningful speedup figure could not be computed.

Finally, using \LNDFS, the total amount of work is decreased, because
workers prune each other's search space. Again, we experimented with
two versions, which are shown in the two right-most columns. We computed
the average runtime of 5 experiments on 16 cores with the
random shared-memory implementation. Note that this is the implementation
that was used in all previous experiments in Section~\ref{s:experiments}.
The figures show again a big improvement over Swarmed \NDFS, even on
a multi-core machine. Interestingly, the fresh successor heuristic also
works very well for the \LNDFS-algorithm, speeding up the algorithm
several times. Similar findings hold for all other parallel \NDFS\ versions
in this paper, because they behave similarly on models with accepting cycles (see Fig.~\ref{f:e_vs_nm} and Fig.~\ref{f:l_vs_n}).


\section{Conclusion}\label{s:conclusion}

In this paper, we experimentally compared two recent parallel
\NDFS-based algorithms, \ENDFS~\cite{Evangelista11} and
\LNDFS~\cite{Laarman11}. We also compared them with Swarmed \NDFS\ and
with the BFS-based algorithm \OWCTYOTF~\cite{owcty-otf}.  We now
summarize the conclusions from our experiments.

For systems with bugs (accepting cycles), both \ENDFS\ and
\LNDFS\ outperform \OWCTYOTF\ by large, so they fully enjoy the
on-the-fly property.  We have also shown that for these cases
\ENDFS\ and \LNDFS\ perform much better than parallel random search,
as in Swarmed \NDFS.

On examples without bugs, it appears that \ENDFS\ beats \LNDFS\ in
most of the cases, due to the fact that there are
still too few red states to prune the blue search in \LNDFS. However,
in a number of other cases \ENDFS\ still scales rather badly,
due to the fact that the sequential repair strategy traverses large
parts of the state space. Interestingly, it is possible to use the
parallel \LNDFS\ algorithm as the repair strategy of \ENDFS.  For this
new combined algorithm, all examples of the \BEEM\ database showed a
decent speedup.

On examples without bugs, \OWCTYOTF\ beats both \LNDFS\ and \ENDFS\ in
a majority of the cases, but still it is slower on a number of other examples.
The combination of \ENDFS\ and \LNDFS, however, provided a good match
for \OWCTYOTF, especially for the larger inputs. This shows that the new
branch of parallel \NDFS\ algorithms is rather promising.

\paragraph{Future work.}
We believe that the last word on parallel LTL model checking has not been said yet.
Although all \NDFS-versions have been implemented in the same framework
so that we compare the algorithmic differences, \OWCTYOTF\ was implemented
in the DiViNE tool. We note that our computation of the $\post$ function
uses the same code from DiViNE. A reimplementation of \OWCTYOTF\ using shared hash
tables will probably increase its speedup, as indicated by results
on pure reachability~\cite{boosting}.

Also the young branch of parallel \NDFS\ algorithms can still be improved.
We have already shown that adding heuristics to direct workers into different regions
of the graph can greatly increase the performance, at least for models with bugs. An interesting question is
if there exists a correct variation on parallel \NDFS\ that can fully share
global information from both the blue and the red search, without the need to
resort to a repair strategy. This would take away the current weak points of
both \ENDFS\ and \LNDFS.

\paragraph{Acknowledgement.}
We thank Ji\v{r}\'\i\ Barnat and Keijo Heljanko for organizing PDMC 2011
and inviting our contribution in this volume. Moreover, we thank Ji\v{r}\'\i\
Barnat for his comments on the time complexity of parallel randomized algorithms,
and Keijo Heljanko for his pointer to the statistical model in~\cite{HyvarinenJN08}.
Finally, we are grateful to Mark Timmer for his help on the statistics.


\bibliographystyle{eptcs}
\bibliography{main}

\begin{thebibliography}{10}
\providecommand{\bibitemdeclare}[2]{}
\providecommand{\urlprefix}{Available at }
\providecommand{\url}[1]{\texttt{#1}}
\providecommand{\href}[2]{\texttt{#2}}
\providecommand{\urlalt}[2]{\href{#1}{#2}}
\providecommand{\doi}[1]{doi:\urlalt{http://dx.doi.org/#1}{#1}}
\providecommand{\bibinfo}[2]{#2}

\bibitemdeclare{book}{principles.mc}
\bibitem{principles.mc}
\bibinfo{author}{C.~Baier} \& \bibinfo{author}{J.P. Katoen}
  (\bibinfo{year}{2008}): \emph{\bibinfo{title}{Principles of Model Checking}}.
\newblock \bibinfo{publisher}{The MIT Press}.

\bibitemdeclare{article}{barnat.brim.multicore}
\bibitem{barnat.brim.multicore}
\bibinfo{author}{J.~Barnat}, \bibinfo{author}{L.~Brim} \&
  \bibinfo{author}{P.~Ro\v{c}kai} (\bibinfo{year}{2010}):
  \emph{\bibinfo{title}{Scalable Shared Memory {LTL} Model Checking}}.
\newblock {\sl \bibinfo{journal}{STTT}}
  \bibinfo{volume}{12}(\bibinfo{number}{2}), pp. \bibinfo{pages}{139--153},
  \doi{10.1007/s10009-010-0136-z}.

\bibitemdeclare{inproceedings}{BBCR10}
\bibitem{BBCR10}
\bibinfo{author}{J.~Barnat}, \bibinfo{author}{L.~Brim},
  \bibinfo{author}{M.~\v{C}e\v{s}ka} \& \bibinfo{author}{P.~Ro\v{c}kai}
  (\bibinfo{year}{2010}): \emph{\bibinfo{title}{{{DiVinE: Parallel Distributed
  Model Checker (Tool paper)}}}}.
\newblock In: {\sl \bibinfo{booktitle}{Parallel and Distributed Methods in
  Verification and High Performance Computational Systems Biology (HiBi/PDMC
  2010)}}, \bibinfo{publisher}{IEEE}, pp. \bibinfo{pages}{4--7},
  \doi{10.1007/978-3-540-88387-6\_20}.

\bibitemdeclare{inproceedings}{owcty-otf}
\bibitem{owcty-otf}
\bibinfo{author}{L.~Barnat}, \bibinfo{author}{L.~Brim} \&
  \bibinfo{author}{P.~Ro{\v c}kai} (\bibinfo{year}{2009}):
  \emph{\bibinfo{title}{A Time-Optimal On-The-Fly Parallel Algorithm for Model
  Checking of Weak {LTL} Properties}}.
\newblock In: {\sl \bibinfo{booktitle}{ICFEM 2009}}, {\sl
  \bibinfo{series}{LNCS}} \bibinfo{volume}{5885}, \bibinfo{publisher}{Springer,
  Heidelberg}, pp. \bibinfo{pages}{407--425},
  \doi{10.1007/978-3-642-10373-5\_21}.

\bibitemdeclare{inproceedings}{BrnoMAP}
\bibitem{BrnoMAP}
\bibinfo{author}{L.~Brim}, \bibinfo{author}{I.~Cern{\'a}},
  \bibinfo{author}{P.~Moravec} \& \bibinfo{author}{J.~Simsa}
  (\bibinfo{year}{2004}): \emph{\bibinfo{title}{Accepting Predecessors Are
  Better than Back Edges in Distributed {LTL} Model-Checking}}.
\newblock In \bibinfo{editor}{A.J. Hu} \& \bibinfo{editor}{A.K. Martin},
  editors: {\sl \bibinfo{booktitle}{FMCAD}}, {\sl \bibinfo{series}{Lecture
  Notes in Computer Science}} \bibinfo{volume}{3312},
  \bibinfo{publisher}{Springer}, pp. \bibinfo{pages}{352--366},
  \doi{10.1007/978-3-540-30494-4\_25}.

\bibitemdeclare{article}{courcoubetis}
\bibitem{courcoubetis}
\bibinfo{author}{C.~Courcoubetis}, \bibinfo{author}{M.Y. Vardi},
  \bibinfo{author}{P.~Wolper} \& \bibinfo{author}{M.~Yannakakis}
  (\bibinfo{year}{1992}): \emph{\bibinfo{title}{{Memory-Efficient Algorithms
  for the Verification of Temporal Properties}}}.
\newblock {\sl \bibinfo{journal}{Formal Methods in System Design}}
  \bibinfo{volume}{1}(\bibinfo{number}{2/3}), pp. \bibinfo{pages}{275--288},
  \doi{10.1007/BFb0023737}.

\bibitemdeclare{inproceedings}{Evangelista11}
\bibitem{Evangelista11}
\bibinfo{author}{S.~Evangelista}, \bibinfo{author}{L.~Petrucci} \&
  \bibinfo{author}{S.~Youcef} (\bibinfo{year}{2011}):
  \emph{\bibinfo{title}{Parallel Nested Depth-First Searches for {LTL} Model
  Checking}}.
\newblock In \bibinfo{editor}{T.~Bultan} \& \bibinfo{editor}{P.-A. Hsiung},
  editors: {\sl \bibinfo{booktitle}{Automated Technology for Verification and
  Analysis 2011}}, {\sl \bibinfo{series}{Lecture Notes in Computer Science}}
  \bibinfo{volume}{6996}, \bibinfo{publisher}{Springer}, pp.
  \bibinfo{pages}{381--396}, \doi{10.1007/978-3-642-24372-1\_27}.

\bibitemdeclare{techreport}{report}
\bibitem{report}
\bibinfo{author}{S.~Evangelista}, \bibinfo{author}{L.~Petrucci} \&
  \bibinfo{author}{S.~Youcef} (\bibinfo{year}{Last accessed 15 Sept 2011}):
  \emph{\bibinfo{title}{Parallel Nested Depth-First Searches for {LTL} Model
  Checking}}.
\newblock \bibinfo{type}{Technical Report}, \bibinfo{institution}{Universit\'e
  Paris 13}.
\newblock
  \urlprefix\url{http://www-lipn.univ-paris13.fr/~evangelista/doc/mc-ndfs.pdf}.

\bibitemdeclare{inproceedings}{gaiser}
\bibitem{gaiser}
\bibinfo{author}{A.~Gaiser} \& \bibinfo{author}{S.~Schwoon}
  (\bibinfo{year}{2009}): \emph{\bibinfo{title}{Comparison of Algorithms for
  Checking Emptiness on B{\"u}chi Automata}}.
\newblock In \bibinfo{editor}{P.~Hlinen{\'y}},
  \bibinfo{editor}{V.~Maty{\'a}{\v{s}}} \& \bibinfo{editor}{T.~Vojnar},
  editors: {\sl \bibinfo{booktitle}{MEMICS'09}}, {\sl
  \bibinfo{series}{OpenAccess Series in Informatics
  (OASIcs)}}~\bibinfo{volume}{13}, \bibinfo{address}{Schloss Dagstuhl,
  Germany}, \doi{10.4230/DROPS.MEMICS.2009.2349}.

\bibitemdeclare{inproceedings}{holzmann-swarm}
\bibitem{holzmann-swarm}
\bibinfo{author}{G.J. Holzmann}, \bibinfo{author}{R.~Joshi} \&
  \bibinfo{author}{A.~Groce} (\bibinfo{year}{2008}):
  \emph{\bibinfo{title}{Swarm Verification}}.
\newblock In: {\sl \bibinfo{booktitle}{ASE}}, \bibinfo{publisher}{IEEE},
  \bibinfo{address}{L'Aquila, Italy}, pp. \bibinfo{pages}{1--6},
  \doi{10.1109/ASE.2008.9}.

\bibitemdeclare{inproceedings}{holzmann-ndfs}
\bibitem{holzmann-ndfs}
\bibinfo{author}{G.J. Holzmann}, \bibinfo{author}{D.~Peled} \&
  \bibinfo{author}{M.~Yannakakis} (\bibinfo{year}{1996}):
  \emph{\bibinfo{title}{{On Nested Depth First Search}}}.
\newblock In: {\sl \bibinfo{booktitle}{The {SPIN} Verification System}},
  \bibinfo{publisher}{American Mathematical Society}, pp.
  \bibinfo{pages}{23--32}.

\bibitemdeclare{inproceedings}{HyvarinenJN08}
\bibitem{HyvarinenJN08}
\bibinfo{author}{A.E.J. Hyv{\"a}rinen}, \bibinfo{author}{T.A. Junttila} \&
  \bibinfo{author}{I.~Niemel{\"a}} (\bibinfo{year}{2008}):
  \emph{\bibinfo{title}{Strategies for Solving {SAT} in Grids by Randomized
  Search}}.
\newblock In \bibinfo{editor}{S.~Autexier}, \bibinfo{editor}{J.~Campbell},
  \bibinfo{editor}{J.~Rubio}, \bibinfo{editor}{V.~Sorge},
  \bibinfo{editor}{M.~Suzuki} \& \bibinfo{editor}{F.~Wiedijk}, editors: {\sl
  \bibinfo{booktitle}{AISC/MKM/Calculemus}}, \bibinfo{series}{LNCS 5144},
  \bibinfo{publisher}{Springer}, pp. \bibinfo{pages}{125--140},
  \doi{10.1007/978-3-540-85110-3\_11}.

\bibitemdeclare{inproceedings}{Laarman11}
\bibitem{Laarman11}
\bibinfo{author}{A.W. Laarman}, \bibinfo{author}{R.~Langerak},
  \bibinfo{author}{J.C. van~de Pol}, \bibinfo{author}{M.~Weber} \&
  \bibinfo{author}{A.~Wijs} (\bibinfo{year}{2011}):
  \emph{\bibinfo{title}{Multi-Core Nested Depth-First Search}}.
\newblock In \bibinfo{editor}{T.~Bultan} \& \bibinfo{editor}{P.-A. Hsiung},
  editors: {\sl \bibinfo{booktitle}{Automated Technology for Verification and
  Analysis 2011}}, {\sl \bibinfo{series}{Lecture Notes in Computer Science}}
  \bibinfo{volume}{6996}, \bibinfo{publisher}{Springer}, pp.
  \bibinfo{pages}{321--335}, \doi{10.1007/978-3-642-24372-1\_23}.

\bibitemdeclare{inproceedings}{freelunch}
\bibitem{freelunch}
\bibinfo{author}{A.W. Laarman}, \bibinfo{author}{J.C. van~de Pol} \&
  \bibinfo{author}{M.~Weber} (\bibinfo{year}{2011}):
  \emph{\bibinfo{title}{Parallel Recursive State Compression for Free}}.
\newblock In \bibinfo{editor}{A.~Groce} \& \bibinfo{editor}{M.~Musuvathi},
  editors: {\sl \bibinfo{booktitle}{SPIN}}, {\sl \bibinfo{series}{Lecture Notes
  in Computer Science}} \bibinfo{volume}{6823}, \bibinfo{publisher}{Springer},
  \bibinfo{address}{Snowbird, USA}, pp. \bibinfo{pages}{38--56},
  \doi{10.1007/978-3-642-22306-8\_4}.

\bibitemdeclare{inproceedings}{boosting}
\bibitem{boosting}
\bibinfo{author}{A.W. {Laarman}}, \bibinfo{author}{J.C. {van de Pol}} \&
  \bibinfo{author}{M.~{Weber}} (\bibinfo{year}{2010}):
  \emph{\bibinfo{title}{{Boosting Multi-Core Reachability Performance with
  Shared Hash Tables}}}.
\newblock In \bibinfo{editor}{N.~Sharygina} \& \bibinfo{editor}{R.~Bloem},
  editors: {\sl \bibinfo{booktitle}{Proceedings of the 10th International
  Conference on Formal Methods in Computer-Aided Design, Lugano, Swiss}},
  \bibinfo{publisher}{IEEE Computer Society}, \bibinfo{address}{USA}, pp.
  \bibinfo{pages}{247--256}.
\newblock \urlprefix\url{http://eprints.eemcs.utwente.nl/18437/}.

\bibitemdeclare{inproceedings}{ltsmin}
\bibitem{ltsmin}
\bibinfo{author}{A.W. {Laarman}}, \bibinfo{author}{J.C. {van de Pol}} \&
  \bibinfo{author}{M.~{Weber}} (\bibinfo{year}{2011}):
  \emph{\bibinfo{title}{Multi-Core {LTSmin}: Marrying Modularity and
  Scalability}}.
\newblock In \bibinfo{editor}{M.~{Bobaru}}, \bibinfo{editor}{K.~{Havelund}},
  \bibinfo{editor}{G.~{Holzmann}} \& \bibinfo{editor}{R.~{Joshi}}, editors:
  {\sl \bibinfo{booktitle}{Proceedings of the Third International Symposium on
  NASA Formal Methods, NFM 2011, Pasadena, CA, USA}}, {\sl
  \bibinfo{series}{LNCS}} \bibinfo{volume}{6617}, \bibinfo{publisher}{Springer
  Verlag}, \bibinfo{address}{Berlin}, pp. \bibinfo{pages}{506--511},
  \doi{10.1007/978-3-642-20398-5\_40}.

\bibitemdeclare{inproceedings}{beem}
\bibitem{beem}
\bibinfo{author}{R.~Pel\'anek} (\bibinfo{year}{2007}):
  \emph{\bibinfo{title}{{BEEM: Benchmarks for Explicit Model Checkers}}}.
\newblock In: {\sl \bibinfo{booktitle}{Proc. of SPIN Workshop}}, {\sl
  \bibinfo{series}{LNCS}} \bibinfo{volume}{4595},
  \bibinfo{publisher}{Springer}, pp. \bibinfo{pages}{263--267},
  \doi{10.1007/978-3-540-73370-6\_17}.

\bibitemdeclare{article}{reif}
\bibitem{reif}
\bibinfo{author}{J.H. Reif} (\bibinfo{year}{1985}):
  \emph{\bibinfo{title}{Depth-first {S}earch is {I}nherently {S}equential}}.
\newblock {\sl \bibinfo{journal}{Information Processing Letters}}
  \bibinfo{volume}{20}(\bibinfo{number}{5}), pp. \bibinfo{pages}{229--234},
  \doi{10.1016/0020-0190(85)90024-9}.

\bibitemdeclare{inproceedings}{schwoon}
\bibitem{schwoon}
\bibinfo{author}{S.~Schwoon} \& \bibinfo{author}{J.~Esparza}
  (\bibinfo{year}{2005}): \emph{\bibinfo{title}{{A Note on On-the-Fly
  Verification Algorithms}}}.
\newblock In \bibinfo{editor}{Nicolas Halbwachs} \& \bibinfo{editor}{Lenore~D.
  Zuck}, editors: {\sl \bibinfo{booktitle}{Tools and Algorithms for the
  Construction and Analysis of Systems}}, {\sl \bibinfo{series}{Lecture Notes
  in Computer Science}} \bibinfo{volume}{3440}, \bibinfo{publisher}{Springer},
  pp. \bibinfo{pages}{174--190}, \doi{10.1007/978-3-540-31980-1\_12}.

\bibitemdeclare{inproceedings}{AutoVerif1986}
\bibitem{AutoVerif1986}
\bibinfo{author}{M.Y. Vardi} \& \bibinfo{author}{P.~Wolper}
  (\bibinfo{year}{1986}): \emph{\bibinfo{title}{An Automata-Theoretic Approach
  to Automatic Program Verification}}.
\newblock In: {\sl \bibinfo{booktitle}{Proc. 1st Symp. on Logic in Computer
  Science}}, \bibinfo{address}{Cambridge}, pp. \bibinfo{pages}{332--344}.
\newblock \urlprefix\url{http://www.cs.rice.edu/~vardi/papers/lics86.pdf.gz}.

\bibitemdeclare{inproceedings}{BrnoOWCTY}
\bibitem{BrnoOWCTY}
\bibinfo{author}{I.~\v{C}ern{\'a}} \& \bibinfo{author}{R.~Pel{\'a}nek}
  (\bibinfo{year}{2003}): \emph{\bibinfo{title}{Distributed Explicit Fair Cycle
  Detection (Set Based Approach)}}.
\newblock In \bibinfo{editor}{T.~Ball} \& \bibinfo{editor}{S.K. Rajamani},
  editors: {\sl \bibinfo{booktitle}{SPIN}}, {\sl \bibinfo{series}{Lecture Notes
  in Computer Science}} \bibinfo{volume}{2648}, \bibinfo{publisher}{Springer},
  pp. \bibinfo{pages}{49--73}, \doi{10.1007/3-540-44829-2\_4}.

\end{thebibliography}

\end{document}